# Electro-Optic Cavities for In-Situ Measurement of Cavity Fields

Michael S. Spencer[1*], Joanna M. Urban[1], Maximilian Frenzel[1], Niclas S. Mueller[1], Olga Minakova[1], Martin Wolf[1], Alexander Paarmann[1], Sebastian F. Maehrlein[1,2,3*]

1. Department of Physical Chemistry, Fritz Haber Institute of the Max Planck Society, 14195 Berlin, Germany
2. Helmholtz-Zentrum Dresden-Rossendorf, Institute of Radiation Physics, 01328 Dresden, Germany
3. Technische Universität Dresden, Institute of Applied Physics, 01062 Dresden, Germany

**Cavity electrodynamics offers a unique avenue for tailoring ground-state material properties, excited-state engineering, and versatile control of quantum matter. Merging these concepts with high-field physics in the terahertz (THz) spectral range opens the door to explore low-energy, field-driven cavity electrodynamics, emerging from fundamental resonances or order parameters. Despite this demand, leveraging the full potential of field-driven material control in cavities is hindered by the lack of direct access to the intra-cavity fields. Here, we demonstrate a new concept of active cavities, consisting of electro-optic Fabry-Pérot resonators, which measure their intra-cavity electric fields on sub-cycle timescales. We thereby demonstrate quantitative retrieval of the cavity modes in amplitude and phase, over a broad THz frequency range. To enable simultaneous intra-cavity sampling alongside excited-state material control, we design a tunable multi-layer cavity, enabling deterministic design of hybrid cavities for polaritonic systems. Our theoretical models reveal the origin of the avoided crossings embedded in the intricate mode dispersion, and will enable fully-switchable polaritonic effects within arbitrary materials hosted by the hybrid cavity. Electro-optic cavities (EOCs) will therefore serve as integrated probes of light-matter interactions across all coupling regimes, laying the foundation for field-resolved intra-cavity quantum electrodynamics.**

*email: maehrlein@fhi-berlin.mpg.de, spencer@fhi-berlin.mpg.de

In recent decades, electromagnetic cavities have been a focus of intense research interest, providing experimental verification of cavity quantum electrodynamics principles, such as Bose-Einstein condensation in condensed matter[1] and atomic systems[2], as well as control over quantum-entangled cavity states[3]. Cavity electrodynamics has even been extended to on-chip photonic cavities[4,5], along with demonstrations of complex cavity-coupled transport[6] and chemistry[7]. Although these developments lead to significant impact across numerous scientific disciplines, they primarily focus on the visible (VIS), infrared (IR), and microwave spectral regions, corresponding to cavity eigenenergies outside the range of the majority of condensed matter excitations. These fundamental, low-energy excitations are driven instead by picosecond (ps) electric field variations, and are therefore native to the $ps^{-1}$ = 1 THz spectral region. Recent advances of carrier-envelope-phase-stable, high-field THz sources – spanning few to tens of THz, with peak field strengths on the order of 1 to 100 MV/cm, respectively[8–10] – open the door to high-field studies in the THz spectral region, thus setting the stage for fundamental investigations of cavity quantum electrodynamics

The objectives of contemporary THz-cavity research are two-fold: Firstly, achieving cavity-induced renormalization of equilibrium material properties[11–14], in the absence of external field driving, and secondly leveraging the resonant interactions between driven cavity modes and an active material, towards ultrafast control and dynamic design of coherently-driven, non-equilibrium states[15]. Cavity-induced renormalization of ground-state material properties has been explored in numerous theoretical works[11,12], and has recently seen experimental verification in a strongly altered insulator-to-metal transition in $1T-TaS_2$[14]. Theoretical proposals include the transition from a quantum paraelectric to ferroelectric ground state in $SrTiO_3$[12], an anti-ferromagnetic to ferromagnetic transition in $\alpha-RuCl_3$[11], and cavity-enhanced electron-electron attractions in pursuit of enhanced superconductivity[13,16,17]. The second class of proposals aim to leverage strongly-coupled light-matter states, termed polaritons, which are formed when the light-matter energy exchange rate exceeds the total system's decoherence rate. Broadband, coherent excitation spanning these new resonances leads to time-domain beating at the Rabi frequency, as observed by coupling cavity modes to either phononic[14,18–21] or magnonic[22–24] resonances. These new states are proposed as promising handles for controlling high-field interactions, such as nonlinear phononics[15], or enhanced phonon-driven superconductivity[25]. Further notable polaritonic effects can be accessed using near-field approaches, for targeting materials with intrinsic polaritons[26,27]. In this work, we focus on bulk cavities, which are readily accessible using far-field radiation, thereby conserving not only momentum, but also complex polarization states of tailored light[28].

Although numerous applications of THz cavity physics have been proposed, the question remains, how can we best exploit the powerful tools of sub-cycle THz physics in the context of cavity electrodynamics? Most importantly, THz time-domain spectroscopy allows for the direct retrieval of amplitude- and phase-resolved electric fields[29]. This technique is particularly well suited to identify polaritonic effects, where strong light-matter coupling is directly observed, evidenced by Rabi oscillations between the polariton eigenstates[18,19,22–24]. Nevertheless, in all THz cavity measurements conducted to date, light-matter coupling has been assessed in an indirect way, by studying light that's emitted from the cavity, i.e. after the fundamental process of light-matter coupling has occurred. THz-based techniques have thus been used as an extension of conventional VIS and near-IR techniques, unlocking amplitude-[19] and phase-resolved[30] cavity mode information. However, these studies suffer from several limitations: the transmitted electric field is significantly weaker compared to the intra-cavity fields, and is distorted after traveling through dispersive cavity mirrors. Furthermore, the mirror substrates introduce reflections[19,31], whose destructive interference can be detrimental to coherent driving. These limitations can be overcome if the electric field is instead measured inside the cavity itself. It is therefore of fundamental scientific interest to enable *in-situ* measurement of cavity electric fields to unlock the full potential of THz cavity electrodynamics, enabling sub-cycle, and local measurements of exotic light-matter coupling states, directly where and when they are emerging within the cavity.

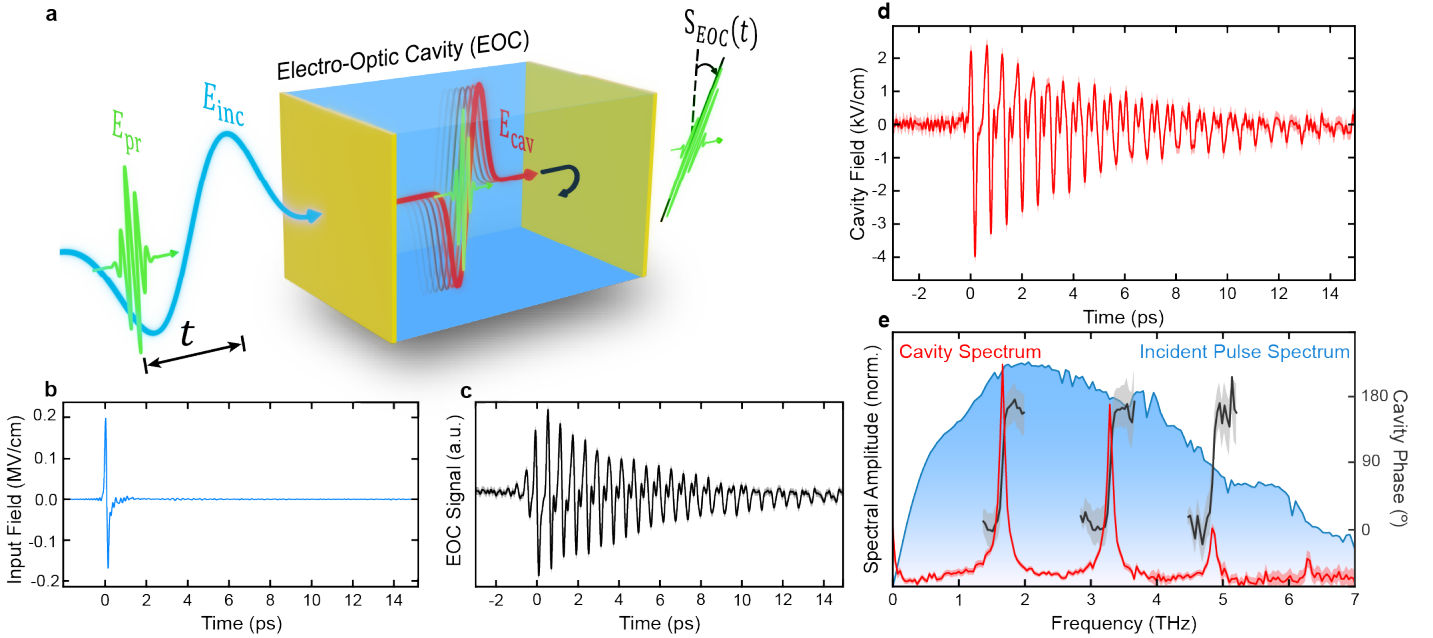

**Fig. 1| Electro-Optic Measurement of Intra-Cavity Fields. a,** The cavity THz electric field (red) induces a local, transient birefringence as it travels inside an α-quartz electro-optic cavity (EOC), which is read out from the transmitted polarization state of an ultrashort co-propagating probing pulse (green), as a function of the relative time delay $t$ with respect to the incident single-cycle THz pulse (blue). **b,** Incident ultra-broadband single-cycle THz field, measured by free-space EO sampling in z-cut α-quartz. **c,** Intra-cavity EO signal generated from the input THz pulse in panel b, for a quartz cavity length of 44 μm, and a nominal deposited gold mirror thickness of 14 nm. The standard error is represented here, and in panels d,e, as a lighter area. **d,** Quantitative, intra-cavity electric field derived from the inverse Fourier transform of the complex cavity spectrum in panel e. **e,** Normalized cavity spectrum derived from the measured EOC signal in panel c, after applying the cavity correction function. Amplitude (red) and frequency-resolved phase (gray) are shown in comparison to the normalized spectrum of the incident broadband pulse in panel b.

In this work, we demonstrate the amplitude- and phase-resolved measurement of intra-cavity fields. For this purpose, we develop a versatile platform of electro-optic cavities (EOCs), by integrating an electro-optic active medium within a Fabry-Pérot cavity. We develop a simple cavity correction function, accounting for both linear and nonlinear dispersive effects, as well as further local and non-local sampling effects, allowing us to extract the quantitative cavity electric fields. We establish EOCs across various cavity lengths and quality factors, and furthermore prototype a continuously-tunable hybrid EOC - a cavity consisting of a pair of electro-optic crystals separated by a tunable air gap. The latter structure surprisingly exhibits avoided crossings of its cavity modes, a typical signature of strongly-coupled oscillators, observed here without the inclusion of an additional active material. To understand this behavior, we develop a cavity field model, and a complementary coupled-oscillator model, which together explain these surprising observations. This detailed understanding directly informs deterministic design of hybrid EOCs, offering in-situ field sampling and tailored material interactions across all light-matter coupling regimes, while operating on sub-cycle time scales.

## Results

### Measuring Intra-Cavity Fields

We demonstrate in **Fig. 1**, to our knowledge, the first direct measurement of the phase-resolved fields inside a Fabry-Pérot cavity. Here, we use the simplest design of an EOC, where the inversion-symmetry broken crystal fully fills the cavity, and its end facets are coated with thin gold films, functioning as the cavity end mirrors. This compact design allows for accurate and sensitive measurement of cavity fields, as depicted schematically in **Fig. 1a**, while simultaneously eliminating complications arising from end-mirror substrates[19,31]. The cavity field is driven by intense, single-cycle THz pump pulses (see **Fig. 1b**), and is probed with synchronized, 20-fs VIS pulses (see Methods). The probing of intra-cavity fields is achieved via the linear electro-optic (EO) effect (Pockels effect), where the THz field induces an effective transient birefringence according to its local amplitude, which is then experienced by the probe pulse. The EO effect is mediated by the 2nd-order nonlinear susceptibility of the inversion-symmetry-broken electro-optic crystal. The resulting transient birefringence is extracted from the transmitted probe pulse polarization via a balanced-detection scheme (Extended Data Fig. 1), as a function of pump-probe delay time $t$, thereby constituting the EOC signal $S_{\text{EOC}}(t)$.

To demonstrate the broad applicability of our approach, we implement the cost-efficient and widely-available crystalline z-cut α-quartz as the cavity medium, which exhibits a relatively weak EO activity[32]. The extremely broadband, single-cycle THz pulse (**Fig. 1b**) which transmits into the cavity will undergo numerous internal reflections, leading to the observed pulse-train signal $S_{\text{EOC}}(t)$ in **Fig. 1c**. This intra-cavity signal, while largely corresponding to the cavity field, nevertheless exhibits clear signatures of probing effects. Most notable are the probe-pulse reflections inside the EOC, evidenced by the clear rise-time of the EOC signal's envelope. To extract the true cavity electric field, we develop a cavity correction function[32–34] to de-convolve the field from the effects of the probing mechanism (see Methods, Extended Data Figure 2). Based on EO linear response theory[32,33], we apply our complex-valued, frequency-domain cavity correction function, $h_{\text{EOC}}(\Omega_{\text{THz}})$, detailed in Methods and Extended Data Fig. 2, to the EOC spectrum via $E_{\text{cav}}(\Omega_{\text{THz}}) = S_{\text{EOC}}(\Omega_{\text{THz}})/h_{\text{EOC}}(\Omega_{\text{THz}})$. In this way, we obtain the cavity electric field (**Fig. 1d**), which is the inverse-Fourier transform of the complex, de-convolved spectrum, $E_{cav}(\Omega_{THz})$ (**Fig. 1e**). Notably, the time-domain electric field no longer displays a rise-time, as the probe pulse reflection effects have been de-convolved from the THz electric field. The cavity spectrum reveals clearly-identified cavity modes (red), with an approximate $\pi$ phase shift (black) across each cavity resonance, in correspondence with theoretical expectations from periodic sampling of an internally-reflected pulse (see Supplementary Information). Thus, we have established reliable extraction of quantitative, phase-resolved intra-cavity electric fields, using a simple cavity correction function - a critical step towards characterizing and controlling field-driven phenomena in more complex cavities.

## Monolithic Electro-Optic Cavities

To engineer cavity fields on demand, we precisely design cavity mirror reflectivities and cavity optical lengths by changing the deposited gold film thickness and quartz crystal length, respectively. In **Figs. 2a,b**, we show the intra-cavity fields, and corresponding spectra, as a function of quartz thickness, thereby adjusting the round-trip time of the THz pulses in the cavity, and therefore the cavity eigenmode spacing. The cavity spectra in **Fig 2b** are offset according to the quartz crystal length, demonstrating perfect agreement with the numerically-calculated cavity eigenfrequencies (light blue lines). To illustrate the capability to systematically tailor cavity quality factors, we show in **Figs. 2c,d** the cavity fields and corresponding spectra as a function of increasing gold film thickness. The cavity fields exhibit an exponential increase in the number of detectable THz pulse internal reflections, and an equivalent linear reduction in the Lorentzian linewidth (see Methods). **Figure 2e** shows the peak fields and quality factors of the 3.3 THz mode (dashed line in **Fig. 2d**), for a systematic study of deposited gold thicknesses. The delayed onset of cavity signatures is attributed to the nucleation of gold islands during the thin-film growth process. These islands must grow large enough to physically overlap for the emergence of macroscopic metallic behavior to occur[35,36]. The experimental data (dots) are in good agreement with our electromagnetic modeling (dashed lines), with which we have experimentally identified metallic behavior onset at $d_{Au} \approx 7.7$ nm (see Methods and Extended Data Fig. 3). This systematic study demonstrates that the cavity quality factor can be tuned to match any fundamental resonance in the THz spectral region and highlights the capability for frequency-tailored EOCs.

## Tunable Hybrid Electro-Optic Cavities

Advancing the concept of frequency-tunable EOCs further, we develop an experimental platform of hybrid EOCs, which allow continuous tuning of the cavity mode frequencies in a single device, while still maintaining the capability for intra-cavity EO sampling. We achieve this, as depicted in **Fig. 3a**, with a pair of quartz crystals ($L_{Qtz}$ = 44 µm), each coated with a gold layer on the exterior facet ($d_{Au}$ = 8 nm) and separated by a piezo motor-controllable air gap of size $L_{Air}$. A representative hybrid EOC signal ($L_{Air}$ = 167 µm), shown in **Fig. 3b**, appears evidently more complicated than those observed for the monolithic cavities shown in **Fig. 2**. The origin of this complexity is revealed in the accompanying spectrum (**Fig. 3c**), which exhibits numerous cavity modes, with non-equidistant mode frequencies, each with differing signal strengths and linewidths. A systematic scan of the air gap lengths in **Figs. 3d,e** unveils continuous evolution of discernable features in the time and frequency domains, respectively. We highlight the avoided crossing signatures in **Fig. 3e** (see details in Extended Data Fig.7c), which typically appear in the presence of strong light-matter coupling, as well as an apparent oscillation of mode strength across the observable frequency range.

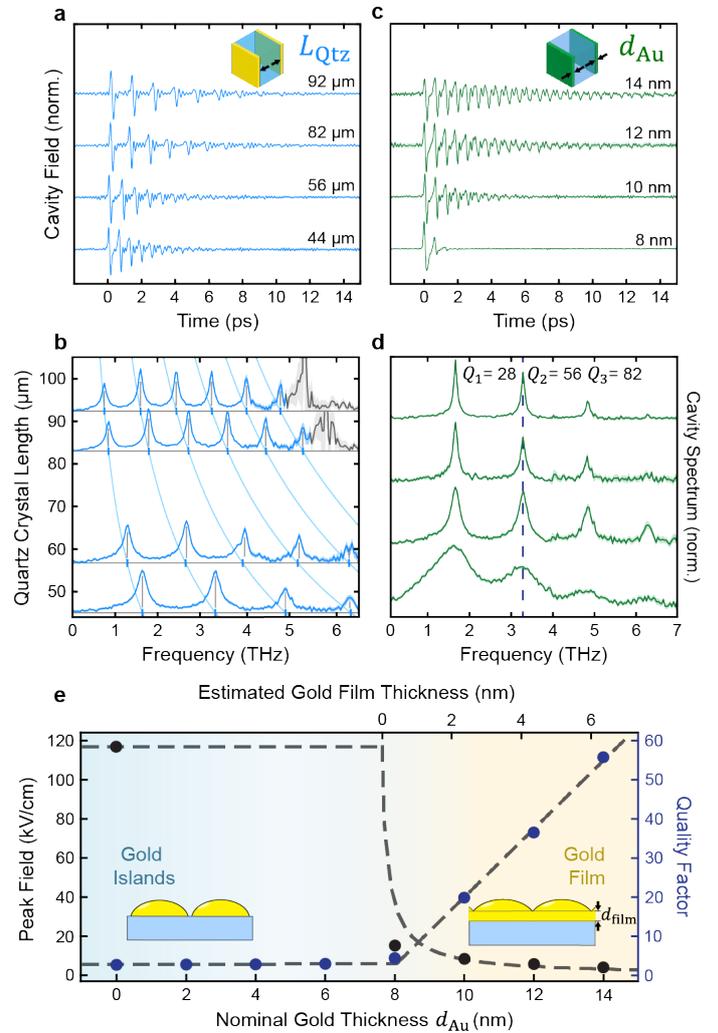

**Fig. 2| Monolithic Electro-Optic Cavity Design. a**, Measured time-domain intra-cavity fields, for quartz cavity lengths of 92, 82, 56, 44 µm. Standard error is represented here, and in panels b-d, as a lighter area. **b**, Cavity spectra corresponding to the cavity fields in panel a, offset according to the cavity length. The intersections (blue ticks) of the computed cavity dispersions (light blue) with the experimental baselines (horizontal gray lines) highlight the perfect agreement to the experimental mode peaks (vertical gray lines). The gray regions are excluded to suppress zero-crossings of the cavity correction function **c,** Measured time-domain intra-cavity fields of a 44µm-long quartz cavity for nominal deposited gold thicknesses of 14, 12, 10, 8 nm. **d**, Cavity spectra corresponding to the fields in panel c, including the mode-specific quality factors for the 14 nm Au mirror cavity (top trace). **e**, Intra-cavity peak fields (black dots) and quality factors (blue dots) of the 3.3 THz mode (dashed line in panel d), as a function of the gold layer thickness. The dashed lines are derived from electromagnetic modelling.

To better understand these experimental features, we simulate using the Scattering Matrix Method (SMM; see Methods) the frequency-resolved transmission of the cavity structure, and can thereafter extract the cavity resonance frequencies, which we overlay onto the experimental spectra in **Fig. 3e**. The perfect agreement of these extracted mode dispersions with our experimental data implies that the observed intensity modulations must arise as a consequence of the hybrid EOC design. Because the SMM simulation does not retain information on the spatial distribution of the fields, it is not capable of providing further physical insight necessary to understand these experimental signatures. Notable among these signatures are the non-equidistant frequency spacing of the modes, and their periodic signal modulation along the horizontal frequency axis in **Fig. 3e,** both necessitating further analysis.

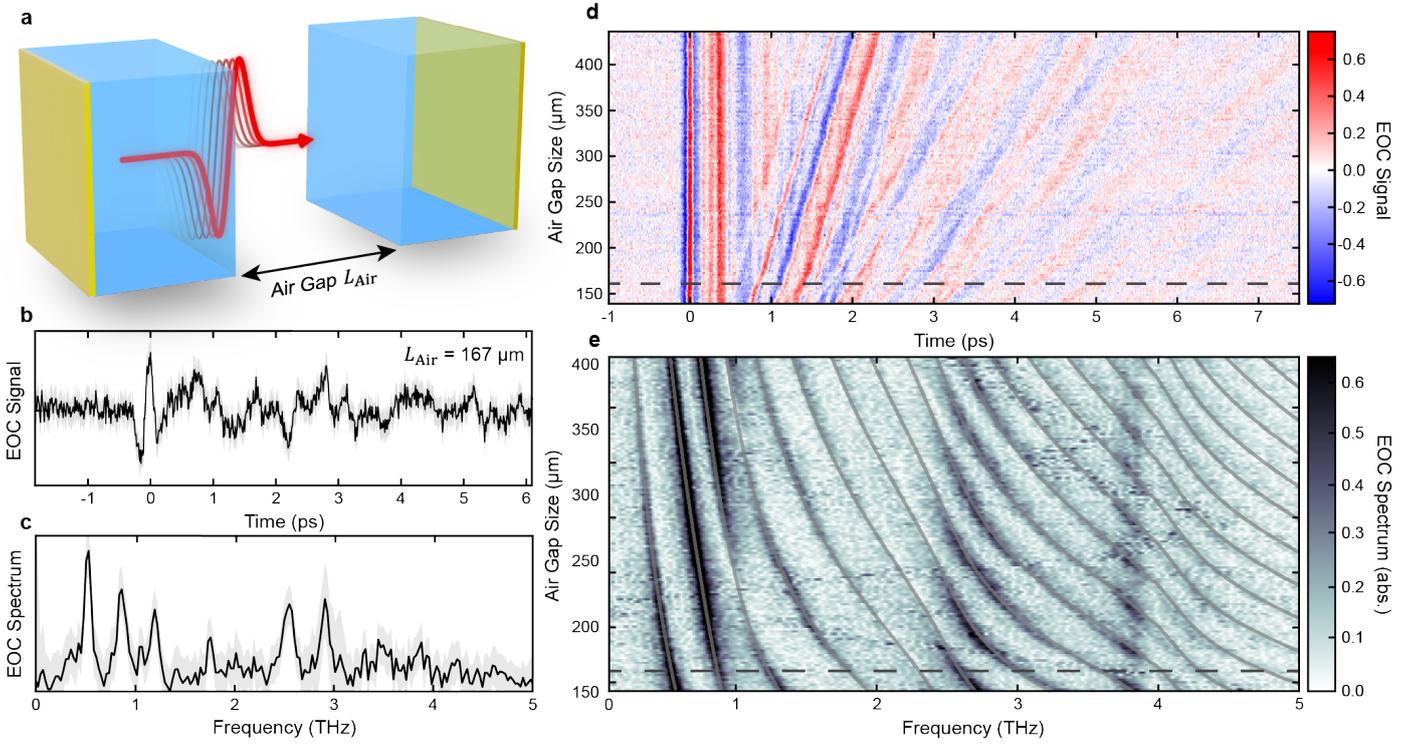

**Fig 3| Tunable Hybrid Electro-Optic Cavities. a**, Tunable hybrid EOC design, implemented by a pair of electro-optic crystals with gold mirrors on the exterior facets, and a remotely-controllable air gap of size $L_{Air}$. **b**, Representative EO signal of a hybrid EOC with an air gap size of $L_{Air}$ = 167 µm. Standard error of EOC signal is represented here and in panel c by light areas. **c,** Corresponding EOC spectrum of panel b. **d**, Systematic measurement of EOC signal (false color) as a function of the air gap size $L_{Air}$, where the dashed line indicates the gap size shown in panels b,c **e**, Continuous EOC spectra (false color), derived from panel d, with mode eigenvalues from Scattering Matrix Method (gray lines) overlaid.

We therefore develop two complementary models – a cavity field-based and a coupled-oscillator model, to understand the signal modulations and the origin of the avoided crossings, respectively. In addition, the cavity-field model[20] provides a quantitative handle to maximize light-matter coupling at the air-quartz interface where samples can be hosted in future studies. To construct this model, we simply identify cavity electric field solutions which are continuous, and whose wavevectors in air and quartz are dictated by the refractive indices (see Methods, and Extended Data Fig. 4). These few simple assumptions yield: (1) the numerically-calculated eigenfrequencies, $\Omega^q(L_{Air})$ for each mode index q, for every choice of air gap size $L_{Air}$, and (2) for each mode the corresponding spatial field distribution $E^q(L_{Air}; z)$. For the coupled-oscillator model, however, we consider the cavity as consisting of interacting sub-cavities: the standing wave modes supported natively in the quartz crystals and in the air gap (see Methods and Extended Data Figs. 6,7). The resulting coupled modes not only perfectly reproduce again the experimental eigenvalues, but also provide an intuitive understanding of the origin of the avoided crossing features. To demonstrate the excellent agreement with the experiment, we overlay the eigenvalues obtained from the cavity-field model onto the nonlinear-susceptibility-corrected experimental EOC spectra in **Fig. 4a** (see also Methods). In the following, we will show how, taken together, these two models provide comprehensive yet intuitive explanations for the experimental signatures.

We identify that the periodic modulation of the EO signal strength emerges as a consequence of the changes in the modes' spatial field distributions across the EOC dispersion. Using the analytical field distributions, we quantify the amplitude coefficient of the field within the quartz layers, normalized against the value inside the air gap (see Methods). Because this parameter directly reflects the amplitude of the mode inside the quartz crystals where we measure the EO signal, we term this ratio the prominence factor $P^q$. By coloring the overlaid eigenfrequencies according to their associated prominence factors in **Fig. 4a**, we highlight the origin of the strong signal modulation arising due to the field suppression within the quartz layers. To further elucidate this relationship, we show in **Fig. 4b** the calculated spatial structure of the eigenmodes' intensities, up to 4 THz, for $L_{Air}$ = 167 µm, where each trace is colored according to the prominence factor. Focusing on the 7th eigenmode's exemplary intensity profile in **Fig. 4c**, we track its evolution with increasing air gap sizes across one period of the prominence factor cycle, for specific cavity configurations denoted by the dots in **Fig. 4a**. Strikingly, the prominence factor is maximized ($P_{max}$ = 1) when the eigenmode has an anti-node at the air-quartz interface, and minimized ($P_{min} = 1/n_{Qtz} \approx 0.5$) when there is a node at the interface. We will demonstrate shortly how this connection between the prominence factor and local field amplitudes will be important for purposes of engineered light-matter interactions.

Beyond the signal amplitude modulation, the cavity-field model also reproduces the EOC dispersion, and avoided-crossing signatures therein. It does not, however, provide an intuitive explanation as to their origin. In the coupled-oscillator model, by contrast, the avoided crossings directly emerge from the coupling between standing-wave resonances within the quartz crystals and those in the air gap (see Methods, and Extended Data Figs. 5-7). Because both models reproduce the EOC dispersion perfectly, the regions of high prominence factor are correlated with eigenmodes possessing a large quartz standing-wave character. This eigenmode character has implications not only for EOC field sensing, but also for future studies of light-matter interactions, as we explore next.

Precise and targeted cavity design will play an important role for future studies of intra-cavity sampling of strong light-matter coupling. As a first step, we demonstrate how the prominence factor should be consulted as a design parameter for light-matter interaction, in addition to its role in EOC sampling sensitivity presented so far. The prominence factor $P^q$ is directly proportional to the electric field at the air quartz interface, according to $E_{int} \propto (P^q - 1/n_{Qtz})/(1 - 1/n_{Qtz})$. Therefore, a thin sample placed or grown on the interior electro-optic crystal facet will experience that same relative interfacial field $E_{int}$. Because the relative light-matter interaction strength scales quadratically with respect to this interfacial field[20], tailoring the frequency-dependent prominence factor $P^q$ plays a key role. We depict in **Fig. 4d** the modeled evolution of the prominence factor and the theoretical signal strength (see Methods) as a function of quartz crystal length and air gap size, for an exemplary target frequency of 2.55 THz, for which the prominence factor of the implemented hybrid EOC is already maximal. With this, we experimentally show in **Fig. 4e** that our hybrid EOC can be actively tuned, from resonant to anti-resonant at a specific design frequency, demonstrating the capability for switchable light-matter coupling that can be directly investigated in an appropriately-tailored EOC. We thus have shown that the cavity-field and coupled-oscillator models described herein provide not only understanding for the measured EOC signal features, but additionally offer critical cavity-design principles for future *in-situ* studies of tunable light-matter coupling.

## Discussion

Our results establish in-situ cavity field measurement in EOCs as a new, integrated probe for low-energy cavity electrodynamics. Time-domain measurements of cavity fields are more direct than conventional intensity-based spectroscopy, where mode details are only implicitly described from linewidths and without phase information[1,37]. Extending field-resolved methods to include intra-cavity sampling now unlocks the additional advantages of enhanced, quantitative measurement of the local cavity fields, in the absence of distortion due to either reflections in cavity substrates, or end-mirror dispersion (see Extended Data Fig. 3). In particular, the reduction in probe field strength due to the cavity reflectivity is very small compared to the relative enhancement of the principal THz pulse peak field strengths inside the cavity, as compared to the transmitted THz pulse, which we quantify as $1/t_{int}$ (see Methods, Extended Data Figs 3a,b). We estimate therefore a relative THz-field enhancement of 25 for the highest quality cavity investigated here.

Our hybrid EOC design extends these concepts by providing continuous tunability and the potential for adding active samples for investigations of intra-cavity light-matter interactions. In the 'empty' hybrid cavity investigated here, we observe a rich mode structure, spurring development of both a field-based model to quantify these cavity modes and their properties, as well as a complementary coupled-oscillator description to gain further understanding of the delicate interplay between the various sub-cavities, which thereafter constitute the hybrid EOC modes. Our detailed analysis of these theoretical vantage points will be highly valuable when considering the addition of an active material, after which the cavity system optical response will become even more intricate. Integration of active materials into hybrid EOCs will yield novel access to light-matter interactions – namely access to energy exchange on sub-Rabi-cycle timescales, and furthermore local probing and even control over tunable light-matter superposition - the latter two unavailable when viewed by conventional cavity transmission techniques. Potential 'active materials' for these in-situ investigations of tunable light-matter interactions include conventional polar semiconductors[38] – oftentimes displaying very large oscillator strengths – , atomically-thin monolayers or heterostructures of transition-metal dichalcogenides[39], hybrid organic-inorganic 3D[21,40] and 2D lead-halide perovskites[41,42], and novel, magnetically-ordered

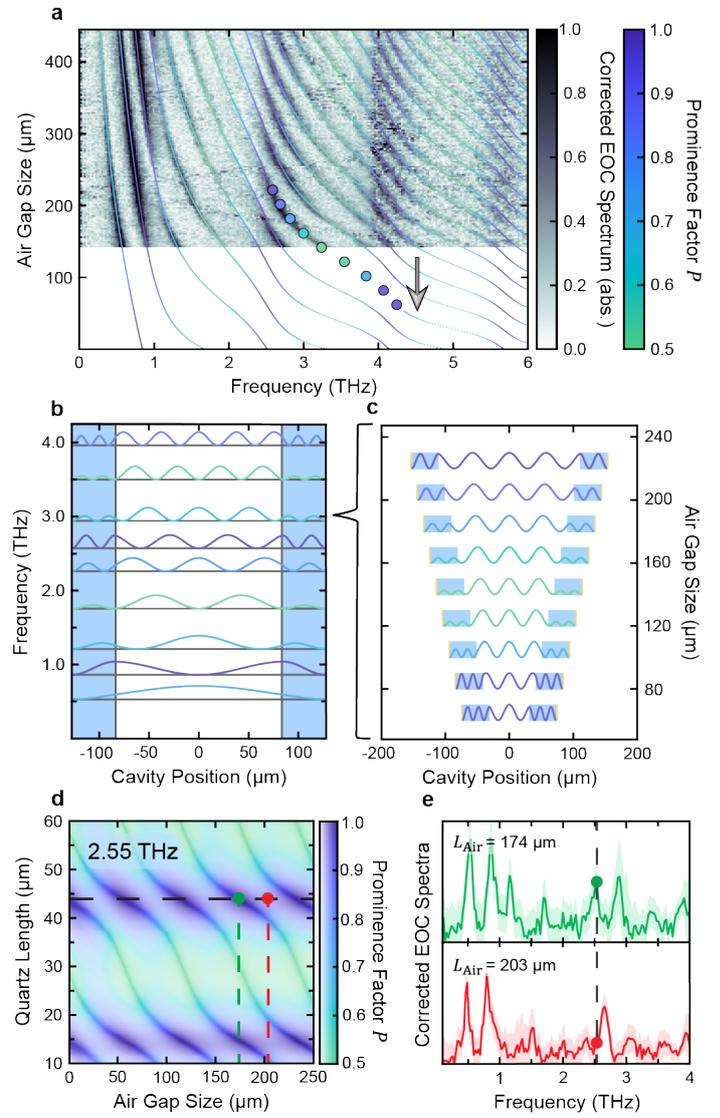

**Fig. 4| Theoretical Modelling of Measured Hybrid Electro-Optic Cavity Features and Active EOC Design Principles. a,** Corrected EOC spectra are displayed (false color), with overlaid eigenvalues obtained from our cavity-field model, color-coded (green→purple) according to their respective prominence factor value $P^q(L_{Air})$. **b,** Spatial mode intensity profiles for gap size $L_{Air}$ = 167 μm, offset by mode frequency, colored according to the prominence factor (green→blue). The blue shaded regions denotes the quartz layers. **c,** Evolution of the 7th cavity mode's intensity profile plotted as a function of air gap length, displaying the spatial origin of the prominence factor (green→blue) oscillation. **d,** The evolution of signal strength (saturation), linewidth, and prominence factor (hue) are shown at 2.55 THz, as a function of quartz length and air gap size. The experimentally-implemented quartz length is denoted by the dashed line. **e,** Corrected EOC spectra at air gap sizes corresponding to a resonance at 2.55 THz (top, green), and anti-resonance (bottom, red), corresponding to those in panel d.

systems[43].

Implementation of EO sampling inside of THz cavities will also significantly advance further areas of contemporary research. As a prominent example, field-resolved probing inside a defined electromagnetic cavity will provide novel opportunities for measurements of electromagnetic vacuum field fluctuations[44,45]. Most notably, a high-quality factor EOC constitutes an advantageous testing ground for measurement of quantum vacuum fluctuations, by efficiently excluding sources of external radiation. Moreover, EOCs are not limited to neither macroscopic environments nor the THz spectral region. Although EO sampling is routinely employed up to the mid-IR spectral region[9], it has recently been extended even into the visible range[46], allowing for future broadband measurements of intra-cavity electric fields. Similar sampling techniques have been used to sample electric fields inside of a metallic antenna-based cavities[47,48], demonstrating that although on-chip photonic implementations lack the dynamic tunability, the general technique is readily implemented in other near-field contexts, including tip-based nano-photonic applications[49]. Additionally, EOCs utilizing quartz are uniquely suited candidates for chiral THz cavity phenomena[50], due to quartz's capability for straightforward and rapid measurement of vectorial electric field trajectories[32].

In conclusion, we have established versatile and compact designs for a new class of active THz cavities, which allow for in-situ retrieval of intra-cavity electric fields. By developing a cavity-correction function formalism for these EOCs, we have demonstrated a rigorous and reliable method to extract absolute fields in a quantitative, and phase-resolved manner. Utilizing straightforward fabrication techniques, we tune the cavities' quality factors and resonance frequencies. Furthermore, we have introduced a hybrid EOC, offering continuously-tunable cavity modes across the entire THz-frequency range, within a single device. This fundamental advancement lays the groundwork for accommodating additional active materials for in-situ measurement of and control over light-matter coupling. We understand the rich hybrid mode structure, including apparent signatures of strong coupling, via a cavity-field model and coupled-oscillator formalism, which will be key to deciphering signatures of light-matter coupling in more complicated devices. Therefore, this work opens new dimensions of THz cavity physics, particularly in the realms of cavity-controlled ground- and excited state material properties. This includes possibilities such as cavity-enhanced THz emission, selectively-driven Floquet states[51], and cavity-controlled nonlinear THz driving[15,52], thus paving the way for comprehensive investigations of THz cavity quantum electrodynamics.

## References

1. Deng, H., Weihs, G., Santori, C., Bloch, J. & Yamamoto, Y. Condensation of Semiconductor Microcavity Exciton Polaritons. *Science* **298**, 199–202 (2002).

2. Brennecke, F. *et al.* Cavity QED with a Bose–Einstein condensate. *Nature* **450**, 268–271 (2007).

3. Mabuchi, H. & Doherty, A. C. Cavity Quantum Electrodynamics: Coherence in Context. *Science* **298**, 1372–1377 (2002).

4. Wallraff, A. *et al.* Strong coupling of a single photon to a superconducting qubit using circuit quantum electrodynamics. *Nature* **431**, 162–167 (2004).

5. Appugliese, F. *et al.* Breakdown of topological protection by cavity vacuum fields in the integer quantum Hall effect. *Science* **375**, 1030–1034 (2022).

6. Paravicini-Bagliani, G. L. *et al.* Magneto-transport controlled by Landau polariton states. *Nat. Phys.* **15**, 186–190 (2019).

7. Garcia-Vidal, F. J., Ciuti, C. & Ebbesen, T. W. Manipulating matter by strong coupling to vacuum fields. *Science* **373**, eabd0336 (2021).

8. Hirori, H., Doi, A., Blanchard, F. & Tanaka, K. Single-cycle terahertz pulses with amplitudes exceeding 1 MV/cm generated by optical rectification in LiNbO3. *Appl. Phys. Lett.* **98**, 091106 (2011).

9. Sell, A., Leitenstorfer, A. & Huber, R. Phase-locked generation and field-resolved detection of widely tunable terahertz pulses with amplitudes exceeding 100 MV/cm. *Opt. Lett.* **33**, 2767 (2008).

10. Rouzegar, R. *et al.* Broadband Spintronic Terahertz Source with Peak Electric Fields Exceeding 1.5 MV/cm. *Phys. Rev. Appl.* **19**, 034018 (2023).

11. Viñas Boström, E., Sriram, A., Claassen, M. & Rubio, A. Controlling the magnetic state of the proximate quantum spin liquid α-RuCl3 with an optical cavity. *NPJ Comput. Mater.* **9**, 202 (2023).

12. Latini, S. *et al.* The ferroelectric photo ground state of SrTiO$_3$: Cavity materials engineering. *Proceedings of the National Academy of Sciences* **118**, e2105618118 (2021).

13. Sentef, M. A., Ruggenthaler, M. & Rubio, A. Cavity quantum-electrodynamical polaritonically enhanced electron-phonon coupling and its influence on superconductivity. *Sci. Adv.* **4**, eaau6969 (2018).

14. Jarc, G. *et al.* Cavity-mediated thermal control of metal-to-insulator transition in 1T-TaS2. *Nature* **622**, 487–492 (2023).

15. Juraschek, D. M., Neuman, T., Flick, J. & Narang, P. Cavity control of nonlinear phononics. *Phys. Rev. Res.* **3**, L032046 (2021).

16. Schlawin, F., Cavalleri, A. & Jaksch, D. Cavity-Mediated Electron-Photon Superconductivity. *Phys. Rev. Lett.* **122**, 133602 (2019).

17. Schlawin, F., Kennes, D. M. & Sentef, M. A. Cavity quantum materials. *Appl. Phys. Rev.* **9**, 011312 (2022).

18. Jarc, G. *et al.* Tunable cryogenic terahertz cavity for strong light–matter coupling in complex materials. *Review of Scientific Instruments* **93**, 033102 (2022).

19. Damari, R. *et al.* Strong coupling of collective intermolecular vibrations in organic materials at terahertz frequencies. *Nat. Commun.* **10**, 3248 (2019).

20. Barra-Burillo, M. *et al.* Microcavity phonon polaritons from the weak to the ultrastrong phonon–photon coupling regime. *Nat. Commun.* **12**, 6206 (2021).

21. Di Virgilio, L. *et al.* Controlling the electro-optic response of a semiconducting perovskite coupled to a phonon-resonant cavity. *Light Sci. Appl.* **12**, 183 (2023).

22. Kritzell, T. E. *et al.* Terahertz Cavity Magnon Polaritons. *Adv. Opt. Mater.* **12,** 2302270 (2023)

23. Baydin, A. *et al.* Magnetically tuned continuous transition from weak to strong coupling in terahertz magnon polaritons. *Phys. Rev. Res.* **5**, L012039 (2023).

24. Blank, T. G. H., Grishunin, K. A. & Kimel, A. V. Magneto-optical detection of terahertz cavity magnon-polaritons in antiferromagnetic HoFeO3. *Appl. Phys. Lett.* **122**, 072402 (2023).

25. Disa, A. S., Nova, T. F. & Cavalleri, A. Engineering crystal structures with light. *Nat. Phys.* **17**, 1087–1092 (2021).

26. Galiffi, E. *et al.* Extreme light confinement and control in low-symmetry phonon-polaritonic crystals. *Nat. Rev. Mater.* **9**, 9–28 (2023).

27. Basov, D. N., Asenjo-Garcia, A., Schuck, P. J., Zhu, X. & Rubio, A. Polariton panorama. *Nanophotonics* **10**, 549–577 (2020).

28. Mitra, S. *et al.* Light-wave-controlled Haldane model in monolayer hexagonal boron nitride. *Nature* **628**, 752–757 (2024).

29. Leitenstorfer, A., Hunsche, S., Shah, J., Nuss, M. C. & Knox, W. H. Detectors and sources for ultrabroadband electro-optic sampling: Experiment and theory. *Appl. Phys. Lett.* **74**, 1516–1518 (1999).

30. Sulzer, P. *et al.* Cavity-enhanced field-resolved spectroscopy. *Nat. Photonics.* **16**, 692–697 (2022).

31. Hazra, S. *et al.* Enhanced Transmission at the Zeroth-Order Mode of a Terahertz Fabry–Perot Cavity. *ACS Omega* **9**, 3000–3005 (2024).

32. Frenzel, M. *et al.* Quartz as an accurate high-field low-cost THz helicity detector. *Optica* **11**, 362 (2024).

33. Gallot, G. & Grischkowsky, D. Electro-optic detection of terahertz radiation. *Journal of the Optical Society of America B* **16**, 1204 (1999).

34. Kampfrath, T., Nötzold, J. & Wolf, M. Sampling of broadband terahertz pulses with thick electro-optic crystals. *Appl. Phys. Lett.* **90**, 231113 (2007).

35. Lee, Y. *et al.* Enhanced terahertz conductivity in ultra-thin gold film deposited onto (3-mercaptopropyl) trimethoxysilane (MPTMS)-coated Si substrates. *Sci. Rep.* **9**, 15025 (2019).

36. Kossoy, A. *et al.* Optical and Structural Properties of Ultra-thin Gold Films. *Adv. Opt. Mater.* **3**, 71–77 (2015).

37. Spencer, M. S. *et al.* Spin-orbit–coupled exciton-polariton condensates in lead halide perovskites. *Sci. Adv.* **7**, eabj7667 (2021).

38. Juraschek, D. M. & Narang, P. Highly Confined Phonon Polaritons in Monolayers of Perovskite Oxides. *Nano. Lett.* **21**, 5098–5104 (2021).

39. Liu, F. *et al.* Disassembling 2D van der Waals crystals into macroscopic monolayers and reassembling into artificial lattices. *Science (1979)* **367**, 903–906 (2020).

40. Frenzel, M. *et al.* Nonlinear terahertz control of the lead halide perovskite lattice. *Sci. Adv.* **9**, aadg3856 (2023).

41. Chen, X. *et al.* Impact of Layer Thickness on the Charge Carrier and Spin Coherence Lifetime in Two-Dimensional Layered Perovskite Single Crystals. *ACS Energy Lett.* **3**, 2273–2279 (2018).

42. Zhang, Z. *et al.* Discovery of enhanced lattice dynamics in a single-layered hybrid perovskite. *Sci. Adv.* **9**, eadg4417 (2023).

43. Mak, K.F., Shan, J. & Ralph, D.C. Probing and controlling magnetic states in 2D layered magnetic materials. *Nat. Rev. Phys.* **1**, 646-661 (2019).

44. Benea-Chelmus, I.-C., Settembrini, F. F., Scalari, G. & Faist, J. Electric field correlation measurements on the electromagnetic vacuum state. *Nature* **568**, 202–206 (2019).

45. Riek, C. *et al.* Direct sampling of electric-field vacuum fluctuations. **350**, 420–423 (2015).

46. Ridente, E. *et al.* Electro-optic characterization of synthesized infrared-visible light fields. *Nat. Commun.* **13**, 1111 (2022).

47. Benea-Chelmus, I.-C. *et al.* Electro-optic interface for ultrasensitive intracavity electric field measurements at microwave and terahertz frequencies. *Optica* **7**, 498 (2020).

48. Kipp, G. *et al.* Cavity electrodynamics of van der Waals heterostructures. Preprint at https://arxiv.org/abs/2403.19745 (2024).

49. Müller, M., Martín Sabanés, N., Kampfrath, T. & Wolf, M. Phase-Resolved Detection of Ultrabroadband THz Pulses inside a Scanning Tunneling Microscope Junction. *ACS Photonics* **7**, 2046–2055 (2020).

50. Hübener, H. *et al.* Engineering quantum materials with chiral optical cavities. *Nat. Mater.* **20**, 438–442 (2021).

51. Hübener, H., De Giovannini, U. & Rubio, A. Phonon Driven Floquet Matter. *Nano. Lett.* **18**, 1535–1542 (2018).

52. Mornhinweg, J. *et al.* Tailored Subcycle Nonlinearities of Ultrastrong Light-Matter Coupling. *Phys. Rev. Lett.* **126**, 177404 (2021).

# Methods

## Experimental Methods

We performed all electro-optic sampling measurements using high-field THz pulses emitted from a large-area spintronic emitter[10], where the pulses were generated by 1kHz, 2.5mJ, 796 nm laser pulses, with Fourier-limited pulse durations of 38fs. The THz pulses at this pump power have maximum field strengths of approximately 200 kV/cm, as measured by EO sampling, and are extremely broadband (maximum spectral strength at 2.0 THz, with a FWHM bandwidth of approximately 3.5 THZ), and temporally short (electric field FWHM of approximately $\tau$ = 100 fs). These THz pulses are sampled using 20-fs pulse duration, 400 pJ, 791 nm broadband laser pulses from a Vitara oscillator.

Cavity electro-optic sampling measurements are performed using a heterodyne detection scheme, where both the transmitted probe pulse and the pulse emitted from the nonlinear polarization are detected simultaneously on a frequency-insensitive photodiode pair. The sum of these two pulses can be treated as a single probe pulse which has experienced an effective field-induced birefringence[33], which is experimentally extracted using a balanced-detection scheme (see Extended Data Fig. 1). This induced rotation is translated into a quantitative cavity electric field using the numerical cavity correction function that is described below.

The gold films were deposited using the standard technique of electron-beam physical vapor deposition. The depositions were performed under ultra-high vacuum conditions of approximately 7 picobars. The nominal deposited gold thicknesses reported herein were monitored via a quartz crystal microbalance.

## Cavity Correction Function

The functional form of the cavity correction function used to analyze the EOC signal (see Supplementary Discussion 3 for an extended discussion) is[32,33]:

$$h_{\text{fun}}(\Omega_{\text{THz}}) = \chi_{\text{eff}}^{(2)}(\Omega_{\text{THz}}) \int_{\omega_{\text{vis}}}^{\Box} \frac{\omega_{\text{vis}}^2}{c_0^2 k(\omega_{\text{vis}})} T_{\text{Pr}}(\omega_{\text{vis}}, \Omega_{\text{THz}}) E_{\text{pr}}^*(\omega_{\text{vis}}) \cdots$$
$$E_{\text{pr}}(\omega_{\text{vis}} - \Omega_{\text{THz}}) G(\omega_{\text{vis}}, \Omega_{\text{THz}}) d\omega_{\text{vis}}. \quad \text{M. 1}$$

Here, $E_{\text{pr}}(\omega_{\text{vis}})$ is the spectrum of the incident visible-frequency probing pulse, and $E_{\text{pr}}(\omega_{\text{vis}} - \Omega_{\text{THz}})$ that of the frequency-shifted pulse emitted from the nonlinear polarization, for a particular THz frequency. The other terms $T_{\text{pr}}$, $\chi_{\text{eff}}^{(2)}$, and $G$ will be discussed in detail shortly. In the assumption of no dispersion within the bandwidth of the probe spectrum, i.e. by assuming that for the quantitative, nonlinear susceptibility $\chi_{\text{eff}}^{(2)}(\Omega_{\text{THz}}, \omega_{\text{vis}}) \approx \chi_{\text{eff}}^{(2)}(\Omega_{\text{THz}})$, the sum-frequency and difference-frequency generation signals differ only in terms of phase, and therefore the total quantitative response function applied to the experimental data is therefore given by:

$$h(\Omega_{\text{THz}}) = \frac{h_{\text{fun}}(\Omega_{\text{THz}}) + h_{\text{fun}}^*(-\Omega_{\text{THz}})}{\int_{\omega_{\text{vis}}}^{\Box} E_{\text{pr}}(\omega_{\text{vis}}) E_{\text{pr}}^*(\omega_{\text{vis}}) d\omega_{\text{vis}}}. \quad \text{M. 2}$$

The cavity correction function accounts for numerous non-local probing effects, each displayed in Extended Data Fig. 2, and discussed now.

The phase mismatch function $G(\omega_{\text{vis}}, \Omega_{\text{THz}})$ (Extended Data Fig. 2a) accounts for the accumulated phase difference between the co-propagating THz and probing fields, and is given by:

$$G(\omega_{\text{vis}}, \Omega_{\text{THz}}; L_{\text{Qtz}}) = \frac{e^{i\Delta k_{\text{co}}(\Omega_{\text{THz}}, \omega_{\text{vis}}) L_{\text{Qtz}}} - 1}{i\Delta k_{\text{co}}(\Omega_{\text{THz}}, \omega_{\text{vis}})}. \quad \text{M. 3}$$

This phase mismatch depends on the fundamental momentum mismatch of the co-propagating THz and visible waves $\Delta k_{\text{co}}(\Omega_{\text{THz}}, \omega_{\text{vis}}) \approx (n_{\text{THz}}(\Omega_{\text{THz}}) - n_{\text{vis}})\Omega_{\text{THz}}$, where we have assumed the visible refractive index is practically dispersionless.

The frequency-dependent nonlinear susceptibility (Extended Data Fig. 2b) is given by:

$$\chi_{\text{eff}}^{(2)}(\Omega_{\text{THz}}) = \chi_e^{(2)} \left[ 1 + \sum_j \frac{C_j \Omega_j^2}{\Omega_j^2 - \Omega_{\text{THz}}^2 - i\Omega_{\text{THz}}\Gamma_j} \right]. \quad \text{M. 4}$$

Here, $\chi_e^{(2)}$ is the purely-electronic (i.e. non-resonant) component of the 2$^{\text{nd}}$-order susceptibility, whose numerical value is taken from Frenzel, et al.[32] The remaining sum represents the resonant ionic terms, written with amplitudes relative to the electronic term. The effective nonlinear susceptibility parameters used here are tabulated for reference in Extended Data Table 2.

The probing pulse transmission $T_{\text{pr}}(\omega_{\text{vis}}, \Omega_{\text{THz}})$ (Extended Data Fig. 2c) is given by the following:

$$T_{\text{Pr}}(\omega_{\text{vis}}, \Omega_{\text{THz}}) = \frac{t_{\text{int}}(\omega_{\text{vis}})}{1 - r_{\text{int}}(\omega_{\text{vis}})^2 e^{i\Omega_{\text{THz}} \tau_{\text{RT}}(\omega_{\text{vis}})}}. \quad \text{M. 5}$$

This serves to account for both the Fabry-Pérot internal reflections (denominator) – separated in time by the visible-frequency cavity round-trip time $\tau_{\text{RT}} = 2n_{\text{Qtz}}(\omega_{\text{vis}})L_{\text{Qtz}}/c_o$, where $c_o$ is the vacuum speed of light– and the transmission out of the cavity $t_{\text{int}}(\omega_{\text{vis}}) \approx 1 - |r_{\text{int}}(\omega_{\text{vis}})|$. The internal reflectivity $r_{\text{int}}$ (defined in Supplementary Discussion 1) is implemented here as a real number, whose correct value is identified by optimal deconvolution with probe cavity reflections, where cavity pulses will otherwise appear (see Extended Data Fig. 2e) as signals at negative delay times. In contrast to the conventional application of a detector response function, we do not consider the Fresnel coefficients for the transfer of the incident pulses into the cavity, as we do not intend to infer the fields incident to the cavity, but strictly the intra-cavity fields.

Finally, for the purpose of comparison with electromagnetic modeling (see Extended Data Figs. 3a, b), we also consider the Fabry-Pérot reflections for the THz intra-cavity pulse (Extended Data Fig. 2d):

$$T_{\text{THz}}(\Omega_{\text{THz}}) = \frac{1}{1 - r_{\text{int}}(\Omega_{\text{THz}})^2 e^{-i\Omega_{\text{THz}} \tau_{\text{RT}}(\Omega_{\text{THz}})}}. \quad \text{M. 6}$$

By including this additional term in the cavity correction function, we identify the principal single-cycle pulse measured inside the cavity (Extended Data Fig. 2g).

We have considered here only measurement of forward-propagation of the principal THz pulse inside the cavity. In Supplementary Discussion 3 we consider both forwards- and backwards propagation of the principle pulse, and characterize our weak measurement of the backwards propagation.

## Monolithic Cavity Characterization

The monolithic cavities are characterized by: (1) round trip times, $\tau_{\text{RT}} = 2n_{\text{Qtz}} L_{\text{Qtz}}/c_o$, and (2) by quality factors, $Q_q = \Omega_q/\delta\Omega_q$, defined for the q$^{\text{th}}$ mode as the ratio between the cavity mode's angular frequency, $\omega_q$, and its linewidth, computed in terms of energy or spectral intensity. There is a common, alternative, definition of the quality factors $Q_q$ in terms of the ratio between the mode frequency and the frequency associated with the energy loss rate: $Q_q = \Omega_q / \left(\frac{l_{\text{RT}}}{\tau_{\text{RT}}}\right)$, where $l_{\text{RT}}$ is the fractional round-trip energy loss. Refer to Supplementary Discussion 2 for further discussion of the round-trip energy-loss rate. Because the cavity frequencies are, neglecting dispersion, multiples of the inverse round trip time, this secondary expression reduces to $Q_q = 2\pi q/l_{\text{RT}}$, where it must however be stressed that this equation is only valid in situations where $l_{\text{RT}} \ll 1$.

To display in **Fig. 2b** the evolution of the cavity eigenfrequencies with quartz crystal length, the cavity eigenfrequencies must be numerically computed if dispersion from the refractive index is to be accounted for. The refractive index used for numerical simulations here is reported in Extended Data Table 1. To achieve this, we identify the frequency $f_{\text{THz}}^q = \Omega_{\text{THz}}^q / 2\pi$ for a given mode index q, and quartz length $L_{\text{Qtz}}$, which satisfy:

$$\frac{\Omega_{\text{THz}}^q}{2\pi} - q \left( \frac{c_o}{2 L_{\text{Qtz}} n_{\text{Qtz}}(\Omega_{\text{THz}}^q)} \right) = 0. \quad \text{M. 7}$$

We also apply in **Fig. 2b** a masking function, in order to suppress the zero-crossing of the phase mismatch function $G(\omega_{\text{vis}}, \Omega_{\text{THz}}; L_{\text{Qtz}})$ in longer quartz crystals (see Supplementary Fig. 6).

In **Fig. 2e**, we show the correspondence between the experimental and simulated fields and quality factors. The experimental field strengths are extracted from the full cavity correction function, where we additionally make a simple correction for probe scattering from gold island nanostructures[36,52]. In this approximation, we track the probe power as a function of gold film thickness, and correct the cavity probe fields by assuming there is a characteristic intensity-proportional scattering associated with the gold islands (see Supplementary Fig. 7). The quality factors are derived from fitting the cavity fields in **Fig. 2d**, according to $Q_q = \omega_q/\delta\omega_q$. The theoretical results are obtained via computation using a three-medium model, where a distinction is made between fields travelling into the cavity, versus out of the cavity (see Supplementary Discussion 1). The results are shown in Extended Data Figs. 3a,b, displaying good agreement not only with the information shown in **Fig. 2e**, but also to the internal THz and visible reflection coefficients inferred from the cavity correction function. In this analysis, we identify best-correspondence Drude model parameters for the gold metal films of $\hbar\omega_{\text{plasma}}$ = 8.5 eV, $\Gamma_{\text{damp}}$ = 8 THz, where we have kept the plasma frequency from Olmon et. al.[53] and varied the damping rate and film onset thickness to identify optimal agreement with the experimentally-measured quality factors and internal cavity maximum field strengths.

This three-medium model is then used to compute the resultant principal pulses emitted from an EOC, using the experimentally-identified Drude parameters, to demonstrate the fundamental effects of dispersion caused by the gold film interfaces (Extended Data Fig. 3, panels c-e). We note that for the relatively thin films used in this study, the effect of spectral dispersion due to the Fresnel coefficients at the gold-quartz interface dominates the pulse dispersion, leading to a suppression of low frequencies in the transmitted pulses. Higher quality cavities, which may be probed with more sensitive electro-optic crystals, will eventually suffer from absorptive effects in thicker gold films, leading to the opposite effect.

## Numerical Simulation – Scattering Matrix Method

To simulate the total field transmission of the hybrid EOC, we implement a numerical solution of Maxwell's Equations called the Scattering Matrix Method (SMM), a variant of the Transfer Matrix Method (TMM), optimized for the case of a strictly isotropic system[54]. TMM propagates the fields through a one-dimensional, layered electromagnetic structure, taking into account the boundary conditions at every interface via a transfer function, and the phase and absorption accumulated across each region. SMM, by contrast, uses a global scattering matrix which is constructed across the entire structure, and used to relate input fields to output fields. This is a more memory-efficient method, and is constructed strictly for linear, homogeneous, and isotropic materials. This efficiency comes at the cost of an inherent lack of spatially-resolved information, due to the formulation of the utilized global scattering matrix.

## Hybrid Cavity – Corrected EOC Spectrum

The 'corrected' EOC spectra presented in **Figs. 4a,e** are obtained by taking the absolute value of the experimental EOC spectra after dividing by the frequency-dependent, second-order, effective susceptibility $\chi_{\text{eff}}^{(2)}(\Omega_{\text{THz}})$, both of which are themselves complex quantities.

$$S_{\text{EOC}}^{\text{corr}}(\Omega_{\text{THz}}) = \left| \frac{S_{\text{EOC}}(\Omega_{\text{THz}})}{\chi_{\text{eff}}^{(2)}(\Omega_{\text{THz}})} \right| \quad \text{M. 8}$$

In this way, we suppress the ionic resonance in $\chi_{\text{eff}}^{(2)}(\Omega_{\text{THz}})$ evident in the raw experimental spectra, in order to highlight the role of the prominence factor in determining the overall signal strength.

## Hybrid Cavity – Simulated Mode Strength

We compute the simulated mode strength $S(\Omega; L_{\text{Air}}, L_{\text{Qtz}})$ displayed in **Fig. 4d,** by constructing a series of complex Lorentzian characteristic lineshapes (see Supplementary Discussion 2), centered at the numerically-identified eigenfrequencies $\Omega_{\text{THz}}^q(L_{\text{Qtz}}, L_{\text{Air}})$, with linewidths $\Gamma = l_{\text{RT}}/\tau_{\text{RT}} P^q(L_{Qtz}, L_{Air})$, i.e. the product of the fundamental loss rate with the prominence factor, divided by the round-trip time. Finally, we multiply by the prominence factor associated to each model oscillator, in order to incorporate to lowest order the effect of probing the field only in the electro-optic crystal.

$$S(\Omega_{\text{THz}}; L_{\text{Air}}, L_{\text{Qtz}}) = \sum_q P^q(L_{\text{Qtz}}, L_{\text{Air}}) \left| \frac{\Gamma/2}{\text{i}(\Omega_{\text{THz}} - \Omega_{\text{THz}}^q(L_{\text{Qtz}}, L_{\text{Air}})) + \Gamma/2} \right| \quad \text{M. 9}$$

## Hybrid Cavity – Field-Based Model

We implement a field-based model to theoretically simulate the cavity eigenfrequencies and eigenmodes that are experimentally observed within the hybrid EOCs. To find these eigenmodes, we look for cavity electric fields which satisfy the following constraints: (1) the field vanishes at the EO crystal-gold boundaries, (2) that the field is continuous across the air-crystal boundaries, and (3) that the first derivative is also continuous across these boundaries. These requirements generate from the initial ansatz of plane waves (see Extended Data Fig. 4a) two classes of solutions, corresponding to modes having either even ($C = D$) or odd ($D = -C$) parity symmetry, and which have cavity eigenfrequencies identified via numerical solutions to the following transcendental equations:

$$\frac{k_{\text{EO}}}{k_o}\cot(k_{\text{EO}} L_{\text{EO}}) = \tan(k_o L_o), \quad \frac{k_{\text{EO}}}{k_o}\cot(k_{\text{EO}} L_{\text{EO}}) = -\cot(k_o L_o), \quad \text{M. 10}$$

where first equation corresponds to even parity symmetry, and the second equation odd. In these equations, the wavevector in air is given by $k_o = \Omega_o/c_o$, where $\Omega_o$ is the THz angular frequency, and $c_o$ the vacuum speed of light. The wavevector inside of the electro-optic crystal is then $k_{\text{EO}} = \sqrt{\varepsilon_{\text{EO}}(\Omega_o)}\Omega_o/c_o$. Finally, $L_{\text{EO}}$ is the length of the electro-optic crystal, and $L_o$ is equal to half of the total air gap $L_{\text{Air}}$, i.e. $L_{\text{Cav}} = 2L_{\text{EO}} + L_{\text{Air}}$. The corresponding electric fields inside the cavity are then determined to be the following:

$$E_{\text{even}}(z) = \begin{cases} \frac{\cos(k_o L_o)}{\sin(k_{\text{EO}} L_{\text{EO}})} \sin\left(k_{\text{EO}}\left(z + \frac{L_{\text{Cav}}}{2}\right)\right), & -L_o - L_{\text{EO}} \leq z \leq -L_o \\ \cos(k_o z), & |z| \leq L_o \\ \frac{-\cos(k_o L_o)}{\sin(k_{\text{EO}} L_{\text{EO}})} \sin\left(k_{\text{EO}}\left(z - \frac{L_{\text{Cav}}}{2}\right)\right), & L_o \leq z \leq L_o + L_{\text{EO}} \end{cases} \quad \text{M. 11}$$

$$E_{\text{odd}}(z) = \begin{cases} \frac{-\text{i}\sin(k_o L_o)}{\sin(k_{\text{EO}} L_{\text{EO}})} \sin\left(k_{\text{EO}}\left(z + \frac{L_{\text{Cav}}}{2}\right)\right), & -L_o - L_{\text{EO}} \leq z \leq -L_o \\ \text{i}\sin(k_o z), & |z| \leq L_o \\ \frac{-\text{i}\sin(k_o L_o)}{\sin(k_{\text{EO}} L_{\text{EO}})} \sin\left(k_{\text{EO}}\left(z - \frac{L_{\text{Cav}}}{2}\right)\right), & L_o \leq z \leq L_o + L_{\text{EO}} \end{cases} \quad \text{M. 12}$$

Further details on obtaining these expressions can be found in Supplementary Discussion 4. The fields for a select few modes are presented in Extended Data Fig. 4d, depicting the effect of air gap size tuning, and its for the prominence factor, as discussed in the next section.

Additionally, we use Snell's law to account for refraction at the air-crystal interfaces, where the incidence angle, $\theta_{\text{inc}}$, and transmitted angle, $\theta_{\text{tr}}$, are defined in terms of the incident wavevector:

$$\sin(\theta_{\text{inc}}) = \frac{k_\perp}{k_{\text{inc}}}; \quad \sin(\theta_{\text{inc}}) = n_{\text{EO}} \sin(\theta_{\text{tr}}). \quad \text{M. 13}$$

Using this, we generalize the field-based model to non-normal angles of incidence, resulting in the updated transcendental equations:

$$\frac{k_{\text{EO}}}{k_o}\cot(k_{\text{EO}}\cos(\theta_{\text{tr}}) L_{\text{EO}}) = \tan(k_o \cos(\theta_{\text{inc}}) L_o); \quad \text{M. 14}$$

$$\frac{k_{\text{EO}}}{k_o}\cot(k_{\text{EO}}\cos(\theta_{\text{tr}}) L_{\text{EO}}) = -\cot(k_o \cos(\theta_{\text{inc}}) L_o). \quad \text{M. 15}$$

We demonstrate in Extended Data Fig. 5 the momentum-dependent dispersion of a representative hybrid cavity's modes, demonstrating why we neglect effects of cavity dispersion in our analysis of hybrid EOC field measurements.

## Hybrid Cavity – Prominence Factor

We define the prominence factor as the absolute value of the normalization coefficient in the EO crystal layer, as identified from the cavity field model:

$$P^q_{\text{even}}(L_{\text{EO}}, L_{\text{Air}}) = \left|\frac{\cos(k_o L_o)}{\sin(k_{\text{EO}} L_{\text{EO}})}\right|; \quad P^q_{\text{odd}}(L_{\text{EO}}, L_{\text{Air}}) = \left|\frac{\sin(k_o L_o)}{\sin(k_{\text{EO}} L_{\text{EO}})}\right|. \qquad \text{M.16}$$

The prominence factor has a detailed evolution for each mode, according to its dispersion (Extended Data Figs. 4b,d). However, when evaluated as a function of frequency, we see that all modes' prominence factors lie on a single curve (see Extended Data Fig. 4c). We identify the functional form of this curve as the following:

$$P(\Omega_{\text{THz}}; L_{\text{EO}}) = \left|\frac{t^{\text{int}}_{\text{EO}}(\Omega_{\text{THz}})}{1 + \left(r^{\text{int}}_{\text{EO}}(\Omega_{\text{THz}})\right)^2 e^{-i\Omega_{\text{THz}}\tau_{\text{RT}}(L_{\text{EO}})}}\right|. \qquad \text{M.17}$$

Here we have used $r^{\text{int}}_{\text{EO}} = (n_{\text{EO}} - 1)/(n_{\text{EO}} + 1)$, and $t^{\text{int}}_{\text{EO}} = 1 - r_{\text{EO}}$ the internal reflection and transmission coefficient from the EO crystal into air, and $\tau_{\text{RT}}(L_{\text{EO}}) = 2n_{\text{EO}} L_{\text{EO}}/c_o$. This function oscillates between the value of 1 and $1/n_{\text{EO}}$ (see Extended Data Fig. 9c), and can be understood as an analog to the Fabry-Pérot transfer functions considered in the cavity correction function, except where the constructive interference conditions are fundamentally modified by the gold mirror, as discussed next.

## Hybrid Cavity – Coupled-Oscillator Model

With this model, we demonstrate that the measured cavity eigenfrequencies may be understood in terms of the interaction of simple standing waves that are sustained in the individual components that constitute the total cavity structure (Extended Data Figs. 6-8). We consider coupling between standing wave resonances in both the electro-optic crystals and the air gap separating them. In the air gap, the modes have path lengths of half-integer multiples of the wavelength, corresponding to typical Fabry-Pérot standing waves (resonances). The modes in the EO crystals have optical path lengths of odd-integer multiples of a quarter of the wavelength in the crystal, differing from the air gap as a consequence of the nodal point imposed by the external gold mirror (Extended Data Fig. 7a). The cavity THz frequencies for these uncoupled modes are therefore given by:

$$f^t_{\text{EO}} = \frac{\Omega^t_{\text{EO}}}{2\pi} = \frac{2t-1}{4}\frac{c_o}{L_{\text{EO}} n_{\text{EO}}(f^t_{\text{EO}})}, \quad t \in \{-\infty, \ldots -1, 1, \ldots, \infty\}, \qquad \text{M.18}$$

$$f^u_{\text{Air}} = \frac{\Omega^u_{\text{Air}}}{2\pi} = \frac{u}{2}\frac{c_o}{L_{\text{Air}}}, \quad u \in \{-\infty, \ldots -1, 0, 1, \ldots, \infty\}. \qquad \text{M.19}$$

For a generic value of THz frequency $f$, we consider a finite number $N_{\text{EO}}$ resonance conditions in the EO crystals and $N_{\text{Air}}$ resonance conditions in the air gap. These correspond to discrete cavity configurations, i.e. values of electro-optic crystal length $L_{\text{EO}}$ and air gap size $L_{\text{Air}}$, respectively, which fulfill:

$$n_{\text{EO}}(f) L^t_{\text{EO}}(f) = \frac{2t-1}{4}\frac{c_o}{f}, \quad t \in \{-N_{\text{EO}}, \ldots -1, 1, \ldots, N_{\text{EO}}\} \equiv T, \qquad \text{M.20}$$

$$L^u_{\text{Air}}(f) = \frac{u}{2}\frac{c_o}{f}, \quad u \in \{-N_{\text{Air}}, \ldots -1, 0, 1, \ldots, N_{\text{Air}}\} \equiv U. \qquad \text{M.21}$$

Normalizing these expressions by the free-space wavelength $\lambda_o = c_o/f$ yields the following series of equations of lines (see Extended Data Figs. 7b,d):

$$\mathcal{M}^t_{\text{EO}} = \frac{2t-1}{4}, \quad \mathcal{M}^u_{\text{Air}} = \frac{u}{2}. \qquad \text{M.22}$$

We identify these quantities generally as the number of optical cycles in the electro-optic medium $\mathcal{M}^t_{\text{EO}} = n_{\text{EO}}(f) L^t_{\text{EO}}(f) f/c_o$, and in air $\mathcal{M}^u_{\text{Air}} = L^u_{\text{Air}}(f) f/c_o$, respectively, and note that both expressions are now completely independent of frequency. We use these quantities as a 'cavity configuration' coordinate system, defining the number of optical cycles in air and quartz to analyze the resultant cavity resonances (see **Fig. 4d** and Extended Data Figs. 7,8) by using the formalism described in the following.

We consider next the coupling of these two families of resonances in our hybrid cavities. We solve, as a first step, for the eigenmodes of a sub-cavity consisting of only the air gap and a single electro-optic crystal, using a coupling block matrix of the form (see Extended Data Figs. 6a, 7):

$$\mathbf{A}_{\text{sub}} = \boldsymbol{\mathcal{M}}^T_{\text{EO}} \oplus \boldsymbol{\mathcal{M}}^U_{\text{Air}} + C\boldsymbol{K}. \qquad \text{M.23}$$

This matrix is written in the direct-sum space of cavity configuration subspaces. That is, we consider the independent variables of $L_{\text{Air}}$ and $L_{\text{EO}}$, and use the coupling matrix $CK$ to identify the coupled-cavity conditions (i.e. eigenvalues $L_{\text{Air}}$ and $L_{\text{EO}}$) that satisfy a particular cavity frequency. The diagonal sub-blocks are constructed using the bare resonances in the subspaces $T$ and U:

$$\boldsymbol{\mathcal{M}}^T_{\text{EO};i,i} = \frac{2T(i)-1}{4}; \quad \boldsymbol{\mathcal{M}}^U_{\text{Air};i,i} = \frac{U(i)}{2}. \qquad \text{M.24}$$

The coupling between these two subspaces is achieved using a single coupling constant $C$, and a matrix $\boldsymbol{K}$ which is filled with the value one in the 'cross space' sub-blocks between $T$ and $U$, and is filled with zeros in the diagonal blocks corresponding to $T$ and $U$ (Extended Data Fig. 6a).

The coupling matrix $\mathbf{A}_{\text{sub}}$ is used to solve for the eigenvectors (i.e. new resonance conditions) in the coupled sub-cavity system, according to (Extended Data Fig. 6b):

$$\mathbf{A}_{\text{sub}} \boldsymbol{a}^l_{\text{sub}} = \mathcal{M}_{\text{sub}} \boldsymbol{a}^l_{\text{sub}}; \quad \mathbf{V}^{-1}_{\text{sub}} \mathbf{A}_{\text{sub}} \mathbf{V}_{\text{sub}} = \boldsymbol{\mathcal{M}}^{T \oplus U}_{\text{sub}}, \qquad \text{M.25}$$

where $\mathbf{V}_{\text{sub}}$ is the matrix of eigenvectors corresponding to $\mathbf{A}_{\text{sub}}$, which executes the change of basis from the uncoupled resonances to the coupled ones.

To obtain the eigenvectors for the entire hybrid cavity, we subsequently couple the sub-cavity eigenvectors to the basis vectors (resonances) in the remaining EO crystal, using the full cavity-configuration space $T_1 \oplus U \oplus T_2$. We consider the full coupling matrix, written as follows:

$$\mathbf{A} = \boldsymbol{\mathcal{M}}^{T_1 \oplus U}_{\text{sub}} \oplus \boldsymbol{\mathcal{M}}^{T_2}_{\text{EO}} + C\boldsymbol{K}_{\text{P}}, \qquad \text{M.26}$$

where the matrix $\boldsymbol{K}_{\text{P}}$ replaces the matrix $\boldsymbol{K}$ used in $\mathbf{A}_{\text{sub}}$ and the values of one are replaced with of the integrated projection of the sub-cavity eigenvectors onto the sub-space of air (Extended Data Fig. 6c), which act as a weighting factor multiplied with the same coupling constant $C$ used in $\mathbf{A}_{\text{sub}}$. The eigenvectors of this full coupling matrix correspond to the physically-allowed air gap sizes and EO crystal lengths which correspond to any given frequency (Ext. Data Figs. 8d-g). We also extract the characters ($\chi_{\text{Air}}, \chi_{\text{EO}}$) of the eigenvectors obtained from $\mathbf{A}$ by analyzing the projection of these eigenvectors onto the relevant subspaces (see Extended Data Fig. 6d, and color-coded eigenvalue dispersions in Ext. Data Figs 7,8).

We identify that it is sufficient to consider eigenvalues only in the range of $\mathcal{M}_{\text{EOC}}$ and $\mathcal{M}_{\text{Air}}$ between 0 and 0.5, constituting an 'irreducible space', thereby taking full advantage of the periodicity in the cavity-configuration space (see Extended Data Figs. 7b,d,f,, 8d,e). Thus, for every desired pair of frequency and EO crystal length, we compute the number of optical cycles in the EO crystal, $\mathcal{M}_{\text{EO}} = L_{\text{EO}} f/c_o$, and identify the unique corresponding eigenvector in the irreducible space. Finally, we identify the value of $\mathcal{M}_{\text{Air}}$ for the identified eigenvector, as well as the higher-order modes which intersect (i.e. $\mathcal{M}_{\text{Air}}$ plus half-integer multiples). In this way, we are able to identify all experimentally-observed cavity eigenfrequencies by investigating only the eigenvectors in the irreducible space. Importantly, the entire cavity dispersion is therefore dictated by a single coupling constant. However, when converted back to an experimentally-relevant basis, e.g. cavity frequency vs. air gap size, the coupling constant will have an inverse-frequency dependence, as is evident in e.g. **Fig. 4a**, and Extended Data Figs. 7g, 8f.

In summary, we have seen that all cavity eigenfrequencies can be understood in terms of a coupling between the simple building blocks which constitute the cavity. Furthermore, all eigenfrequencies are derived from an entirely frequency-independent irreducible space, determined via a single coupling constant (neglecting dispersion), thus representing the most compact representation of the physics demonstrated by the hybrid EOC.

## Hybrid Cavity - Effects of Refractive Index

Through a systematic study of the effect of refractive index using the field-based and coupled-oscillator models, we identify that there is a strong trade-off for using more sensitive EO crystals, such as zinc telluride (ZnTe) or gallium phosphide (GaP). We observe that the interaction strength in the coupled-oscillator model scales roughly with the transmission coefficient between air and the EO crystal (Extended Data Fig. 9c). Thus, for higher refractive indices, the frequency range in which air-crystal coupling occurs noticeably shrinks, leading to significantly more prominent features in cavity dispersion when tuning the air gap size (see Extended Data Figs. 7g, 8f, & 9d).

Because strong coupling is ideally studied when a resonance of interest is energetically aligned with an EO crystal resonance, we identify quartz as the ideal electro-optic platform of choice for hybrid EOC studies of light-matter interaction, as the large refractive indices of ZnTe and GaP are likely to dominate the hybrid cavity electro-optical features.

## Data Availability

All data used to support the conclusions in this manuscript are present in this text and in the Supplementary Information. The experimental data will be made available at a public online repository after publication of this manuscript.

## References

51. Yuen-Ron Shen. *The Principles of Nonlinear Optics*. (Wiley, 1984).
52. Axelevitch, A., Apter, B. & Golan, G. Simulation and experimental investigation of optical transparency in gold island films. *Opt. Express* **21**, 4126 (2013).
53. Olmon, R. L. *et al.* Optical dielectric function of gold. *Phys. Rev. B* **86**, 235147 (2012).
54. Rumpf, R. C. IMPROVED FORMULATION OF SCATTERING MATRICES FOR SEMI-ANALYTICAL METHODS THAT IS CONSISTENT WITH CONVENTION. *Progress In Electromagnetics Research B* **35**, 241–261 (2011).

## Acknowledgements

The authors thank Sven Kubala for deposition of the gold films. S.F.M. acknowledges support and funding from the Emmy Noether Programme of the Deutsche Forschungsgemeinschaft (469405347).

## Author contributions

M.S.S. and S.F.M. conceived the experimental concept and its design. All experiments were performed by M.S.S. Data analysis was performed by M.S.S. with support from J.M.U., and M.F. The cavity correction function was developed by M.S.S. and M.F., the cavity field model by M.S.S., O.M., and J.M.U, and the oscillator model by M.S.S., N.S.M., and J.M.U. The paper was written by M.S.S., A.P., and S.F.M., with input, discussion, and analysis from all coauthors.

## Competing Interests

The authors declare that they have no competing interests.

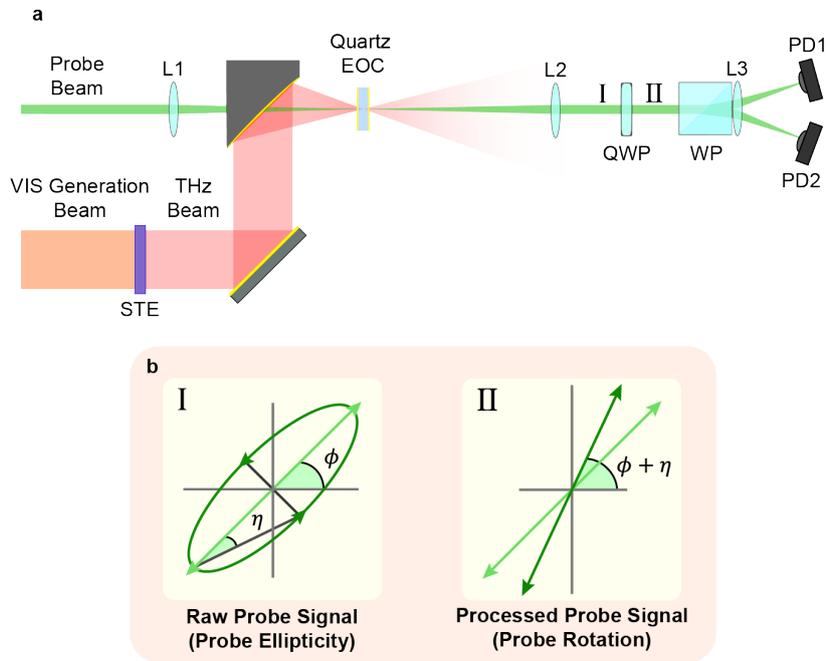

**Extended Data Fig. 1| Balanced Detection of Electro-Optic Cavity Signal. a,** The high-power visible pulse pumps the spintronic emitter, generating a THz pulse, which is subsequently focused into the quartz EOC. The THz pulse reflects periodically inside the cavity, and travels alongside the probe pulse co-focused into the cavity, thereby generating EOC signal. This signal results in an effective ellipticity imparted to the probe beam pulse, which is measured using the balanced detection optics. **b**, The polarization of the probe beam after interaction in the cavity is plotted at the various stages in the balanced detection scheme. The gray lines denote the projection of the polarization state onto the x and y axes, and the lighter green vector denotes the polarization state of the probe in the absence of any THz field interaction. After THz field interaction in the EOC, the probe acquires an ellipticity, $\eta$. This ellipticity is converted into a rotation of the linear polarization angle by use of a QWP. This rotation is measured by splitting the polarization into vertical and horizontal components using the WP and measuring the difference signal from the intensities on a photodiode pair.

**Figure Legend:** L1, L2, L3 – Lenses 1-3, STE – Spintronic Emitter EOC – Electro-optic Cavity, QWP – Quarter Waveplate, WP – Wollaston Prism, PD1, PD2- Photodiodes 1,2

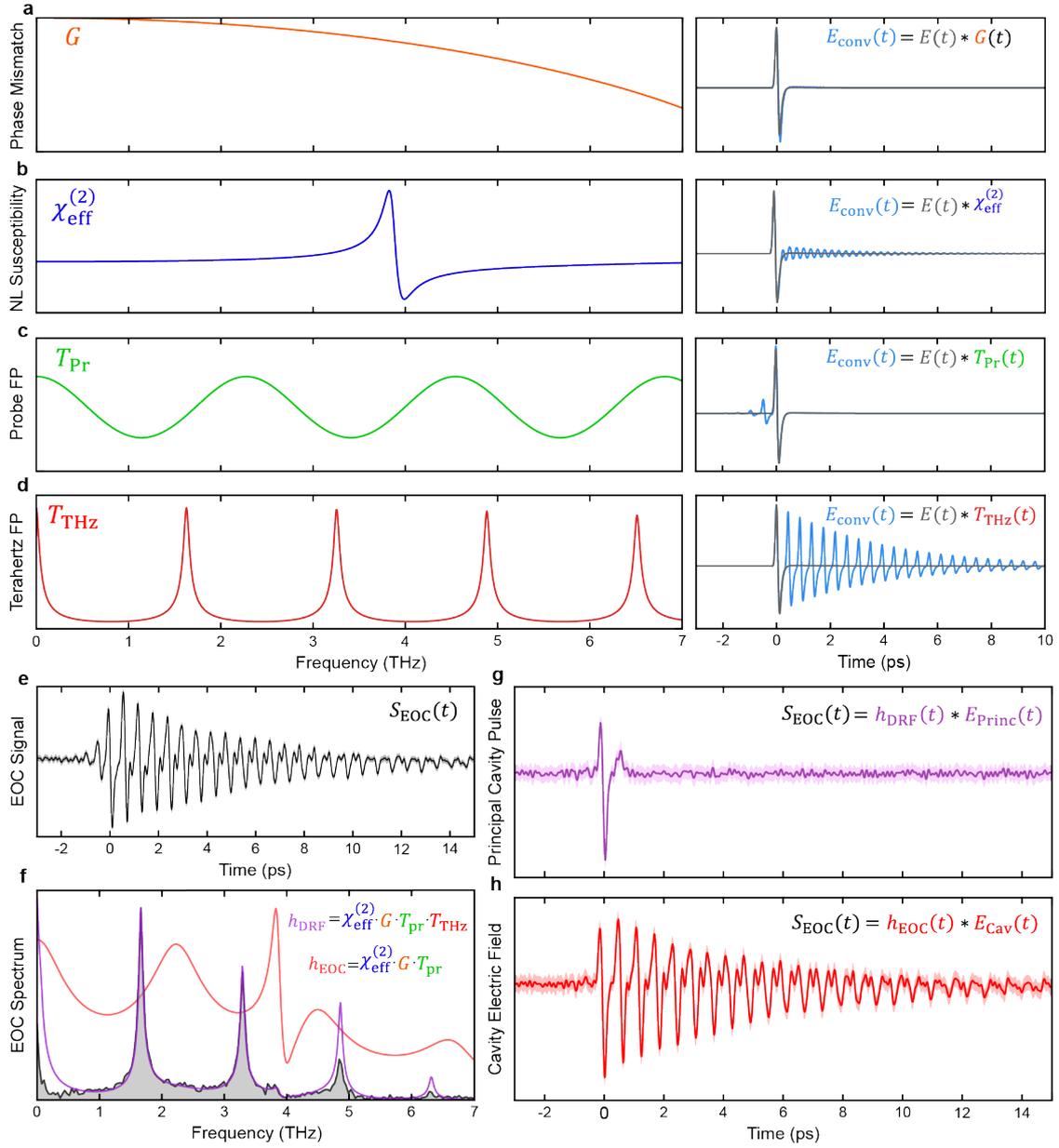

**Extended Data Fig. 2 | Cavity Correction Function.** The most important frequency-dependent components of the cavity correction function are shown as a function of terahertz frequency, for the EOC conditions investigated in **Fig. 1** For panels a-d, the left panel shows the absolute value of the function in the frequency domain, and the right side the result of convolution with a theoretical principal pulse. **a,** The absolute value of the frequency-dependent phase-mismatch (orange). **b,** The absolute value of the frequency-dependent effective 2$^{nd}$ order nonlinear susceptibility (blue). **c,** The absolute value of the frequency-dependent effective 2$^{nd}$ order nonlinear susceptibility (blue). **d,** The absolute value of the frequency-dependent cavity fields transfer function at probe frequencies (green). **e,** The absolute value of the frequency-dependent cavity fields transfer functio for THz frequencies (red). For panels a-d, plotted additionally on the right are the convolution of these (complex) functions with a model THz pulse. The nonlinear susceptibility displays the signature of a phonon resonance, the phase-mismatch displays a slight broadening of the THz pulse (more severe for thicker EOS crystals), the cavity field for the probe shows a negative-time pulse train (very low Q), and the cavity field for the THz field shows a positive-time pulse train (higher Q). **e,** Raw Experimental EOC signal, as displayed in **Fig 1**. **f,** Raw Experimental EOC spectrum (black), EOC correction function (red), and cavity detector response function (purple). The detector response function differs from the correction function by the additional term for de-convolving the Fabry-Pérot resonances for the intra-cavity THz pulse. **g,** The result of deconvolving the EOC signal with the cavity detector response function yields the principal single-cycle pulse, demonstrating appropriate choice of THz internal cavity reflectivity. **h,** Cavity electric field, resulting from deconvolving the EOC signal with the cavity correction function, as plotted in **Fig. 1**.

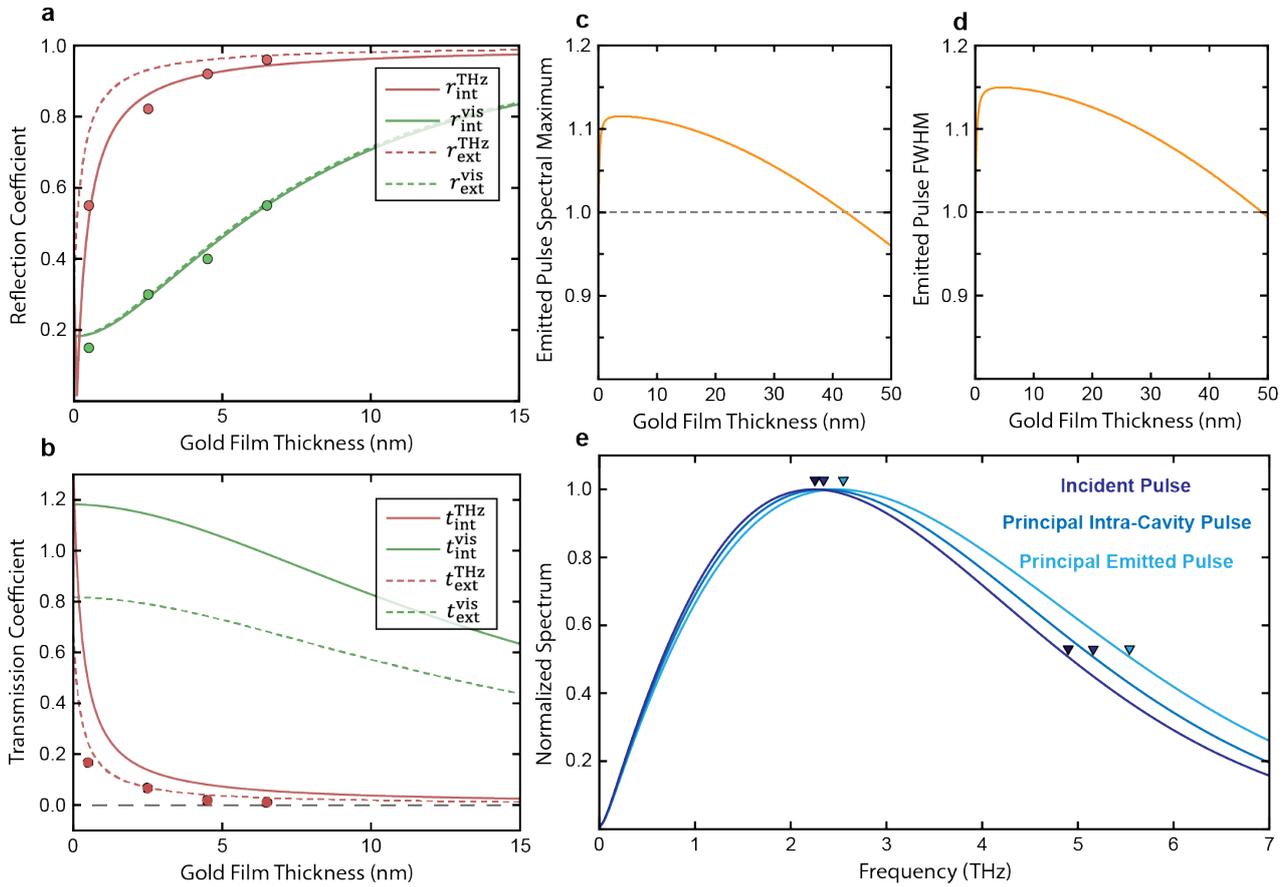

**Extended Data Fig. 3| Gold Film Mirror Optical Properties, & Dispersion. a,b,** Field transmission and reflection coefficients for THz and visible frequencies (0.35 THz and 380 THz, here, respectively) calculated for both internal and external pulses, as defined in Supplementary information. Dots represent experimentally-inferred data, along with the results shown in **Fig. 3e**. The reflection values are identified via inspection of deconvolved cavity fields, and the transmission values from the resulting cavity peaks fields, after treatment using the numerical cavity correction function. All experimental reflectivity values are identified independently of the thin-film electromagnetic model. **c,d,** Using a theoretical THz pulse and the electromagnetic three-medium model, we compute emitted pulses for various different gold film thicknesses. In panel c we show the frequency of the emitted pulse's strongest spectral weight. In panel d, we show extracted FWHM of the emitted pulse. **e,** The normalized spectrum of the incident pulse, principal intra-cavity pulse, and principal emitted pulse, for $d_{\mathrm{Au}}$ = 5 nm. The spectral maximum and FWHM frequency are denoted for each pulse with a colored triangle.

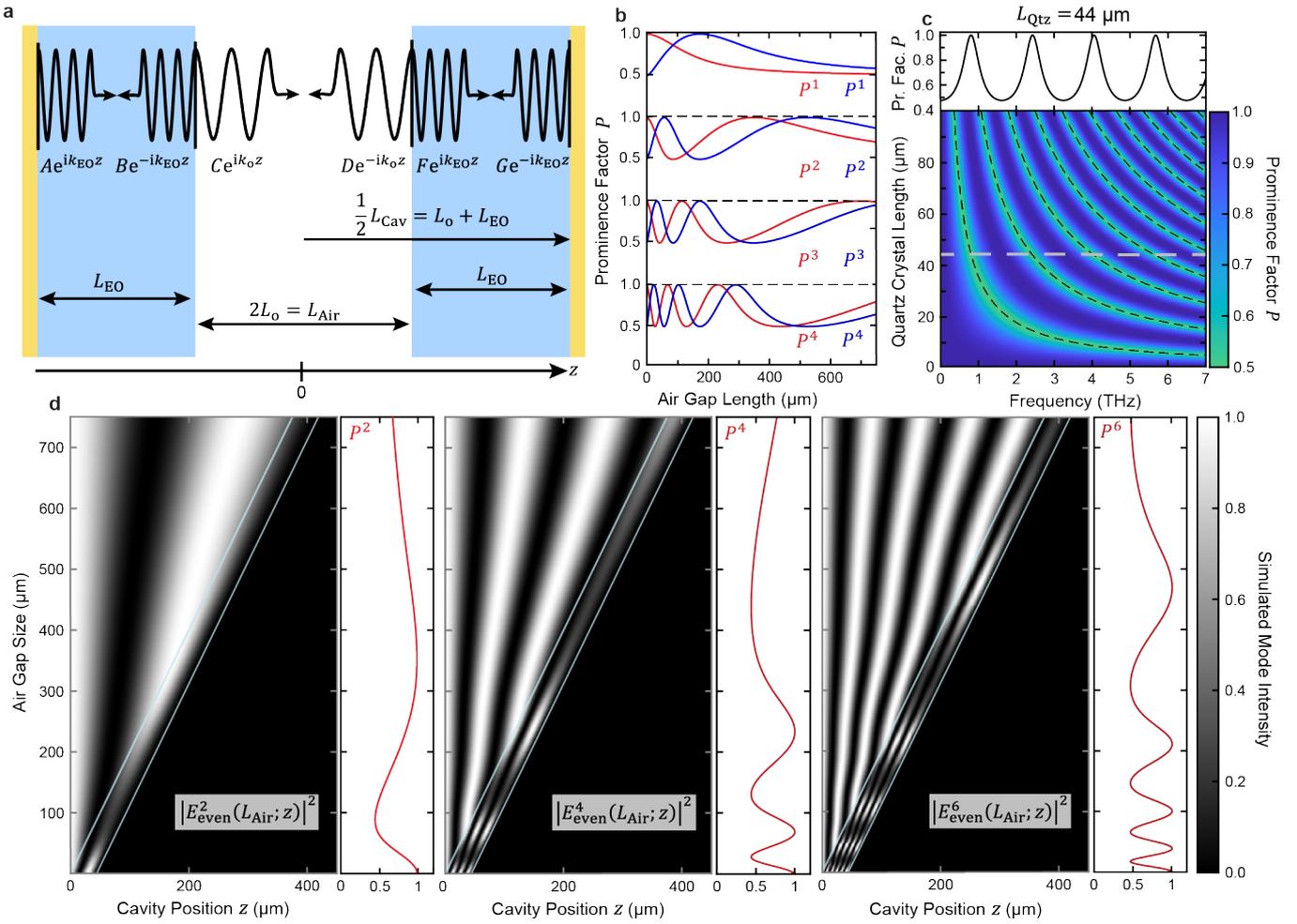

**Extended Data Fig. 4| Field Model. a,** Formulation of the hybrid EOC cavity-field model. In each section there is a forward- and backwards-propagating planewave. The wavevector in the electro-optic crystal and air is dictated by the refractive index at the corresponding frequency in either layer. **b,** The prominence factor, depicted for the first 4 even (red) and odd (blue) modes, as a function of the air gap length, simulated for $n_{EO} = 2$. **c,** The prominence factor is plotted as a function of the quartz crystal length (lower panel), and for the specific quartz thickness experimentally implemented here $L_{Qtz} = 44$ μm (upper panel). Both panels depict the prominence factor in the absence of dispersion, using $n_{EO} = n_{Qtz} = 2.1$. **d,** Intensity profiles of the 2$^{nd}$, 4$^{th}$, and 6$^{th}$ even cavity modes, as a function of the air gap length. Only positive spatial positions are displayed, as the intensity for all modes has even mirror symmetry. The boundaries of the quartz are depicted in all plots with the light blue lines. For each intensity plot, the associated integrated intensity ratio and prominence factors are displayed to the right side of the false-color plot, which demonstrate local maxima wherever the amplitude in quartz is maximized.

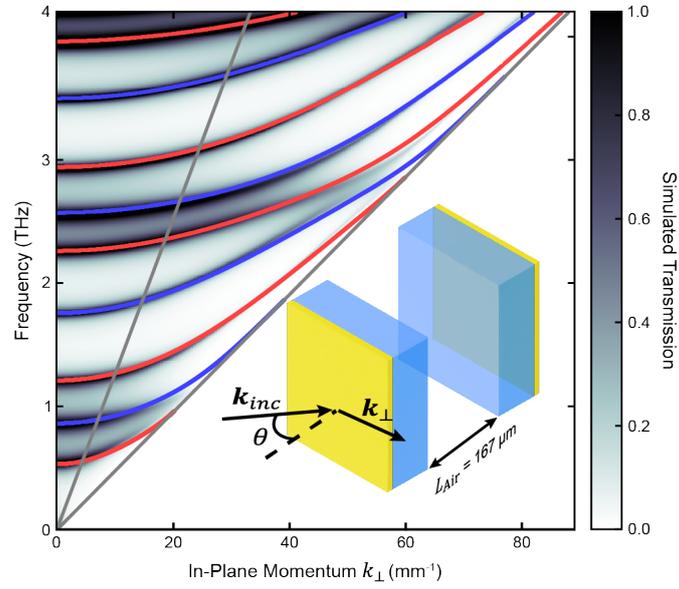

**Extended Data Fig. 5| Cavity Momentum Dispersion.** Depiction of the non-normal incidence generalized cavity field model eigenvalues, colored red (even) and blue (odd) in reference to the parity symmetry. These eigenvalues are overlaid onto SMM-simulated transmittance. Both simulations are computed for $L_{Qtz}$ = 44 μm and $L_{Air}$ = 167 μm. We depict the light line corresponding to grazing incidences (exterior light line), and corresponding to an incidence angle of 22.5° (interior light line). Based on the beam size and focal length of the parabolic mirror, the majority of our THz radiation lies within this interior light line. We observe that there is minimal dispersion of the modes within this region, thus indicating that momentum dispersion doesn't play a large role in our hybrid EOC measurements.

**a**

$$\mathbf{A}_{\text{sub}} = \mathcal{M}_{\text{EO}}^T \oplus \mathcal{M}_{\text{Air}}^U + CK$$

$$\mathbf{A}_{\text{sub}} = \begin{pmatrix} \mathcal{M}_{\text{EO}}^T & \mathbb{0} \\ \mathbb{0} & \mathcal{M}_{\text{Air}}^U \end{pmatrix} + C \begin{pmatrix} \mathbb{0} & \mathbb{1} \\ \mathbb{1} & \mathbb{0} \end{pmatrix} = \begin{pmatrix} \mathcal{M}_{\text{EO}}^{-N_{\text{EO}}} & 0 & 0 & 0 & C & C & C & C & C \\ 0 & \mathcal{M}_{\text{EO}}^{-N_{\text{EO}}+1} & 0 & 0 & C & C & C & C & C \\ 0 & 0 & \mathcal{M}_{\text{EO}}^{N_{\text{EO}}-1} & 0 & C & C & C & C & C \\ 0 & 0 & 0 & \mathcal{M}_{\text{EO}}^{N_{\text{EO}}} & C & C & C & C & C \\ C & C & C & C & \mathcal{M}_{\text{Air}}^{-N_{\text{Air}}} & 0 & 0 & 0 & 0 \\ C & C & C & C & 0 & \mathcal{M}_{\text{Air}}^{-N_{\text{Air}}+1} & 0 & 0 & 0 \\ C & C & C & C & 0 & 0 & \mathcal{M}_{\text{Air}}^{0} & 0 & 0 \\ C & C & C & C & 0 & 0 & 0 & \mathcal{M}_{\text{Air}}^{N_{\text{Air}}-1} & 0 \\ C & C & C & C & 0 & 0 & 0 & 0 & \mathcal{M}_{\text{Air}}^{N_{\text{Air}}} \end{pmatrix}$$

**b**

$$\mathbf{V}_{\text{sub}}^{-1} \mathbf{A}_{\text{sub}} \mathbf{V}_{\text{sub}} = \mathcal{M}_{\text{sub}}^{T \oplus U} \qquad \mathbf{P}_{\text{Air}} = \mathbb{0}^T \oplus \mathbb{1}^U = \begin{pmatrix} \mathbb{0} & \mathbb{0} \\ \mathbb{0} & \mathbb{1} \end{pmatrix}$$

$$\mathbf{V}_{\text{sub}} = \begin{pmatrix} c_1^{-N_{\text{EO}}} & c_2^{-N_{\text{EO}}} & c_3^{-N_{\text{EO}}} & c_4^{-N_{\text{EO}}} & c_5^{-N_{\text{EO}}} & c_6^{-N_{\text{EO}}} & c_7^{-N_{\text{EO}}} & c_8^{-N_{\text{EO}}} & c_9^{-N_{\text{EO}}} \\ c_1^{-N_{\text{EO}}+1} & c_2^{-N_{\text{EO}}+1} & c_3^{-N_{\text{EO}}+1} & c_4^{-N_{\text{EO}}+1} & c_5^{-N_{\text{EO}}+1} & c_6^{-N_{\text{EO}}+1} & c_7^{-N_{\text{EO}}+1} & c_8^{-N_{\text{EO}}+1} & c_9^{-N_{\text{EO}}+1} \\ c_1^{N_{\text{EO}}-1} & c_2^{N_{\text{EO}}-1} & c_3^{N_{\text{EO}}-1} & c_4^{N_{\text{EO}}-1} & c_5^{N_{\text{EO}}-1} & c_6^{N_{\text{EO}}-1} & c_7^{N_{\text{EO}}-1} & c_8^{N_{\text{EO}}-1} & c_9^{N_{\text{EO}}-1} \\ c_1^{N_{\text{EO}}} & c_2^{N_{\text{EO}}} & c_3^{N_{\text{EO}}} & c_4^{N_{\text{EO}}} & c_5^{N_{\text{EO}}} & c_6^{N_{\text{EO}}} & c_7^{N_{\text{EO}}} & c_8^{N_{\text{EO}}} & c_9^{N_{\text{EO}}} \\ c_1^{-N_{\text{Air}}} & c_2^{-N_{\text{Air}}} & c_3^{-N_{\text{Air}}} & c_4^{-N_{\text{Air}}} & c_5^{-N_{\text{Air}}} & c_6^{-N_{\text{Air}}} & c_7^{-N_{\text{Air}}} & c_8^{-N_{\text{Air}}} & c_9^{-N_{\text{Air}}} \\ c_1^{-N_{\text{Air}}+1} & c_2^{-N_{\text{Air}}+1} & c_3^{-N_{\text{Air}}+1} & c_4^{-N_{\text{Air}}+1} & c_5^{-N_{\text{Air}}+1} & c_6^{-N_{\text{Air}}+1} & c_7^{-N_{\text{Air}}+1} & c_8^{-N_{\text{Air}}+1} & c_9^{-N_{\text{Air}}+1} \\ c_1^{0} & c_2^{0} & c_3^{0} & c_4^{0} & c_5^{0} & c_6^{0} & c_7^{0} & c_8^{0} & c_9^{0} \\ c_1^{N_{\text{Air}}-1} & c_2^{N_{\text{Air}}-1} & c_3^{N_{\text{Air}}-1} & c_4^{N_{\text{Air}}-1} & c_5^{N_{\text{Air}}-1} & c_6^{N_{\text{Air}}-1} & c_7^{N_{\text{Air}}-1} & c_8^{N_{\text{Air}}-1} & c_9^{N_{\text{Air}}-1} \\ c_1^{N_{\text{Air}}} & c_2^{N_{\text{Air}}} & c_3^{N_{\text{Air}}} & c_4^{N_{\text{Air}}} & c_5^{N_{\text{Air}}} & c_6^{N_{\text{Air}}} & c_7^{N_{\text{Air}}} & c_8^{N_{\text{Air}}} & c_9^{N_{\text{Air}}} \end{pmatrix}$$

$$\mathbf{P}_{\text{Air}} \mathbf{V}_{\text{sub}} = \begin{pmatrix} 0 & 0 & 0 & 0 & 0 & 0 & 0 & 0 & 0 \\ 0 & 0 & 0 & 0 & 0 & 0 & 0 & 0 & 0 \\ 0 & 0 & 0 & 0 & 0 & 0 & 0 & 0 & 0 \\ 0 & 0 & 0 & 0 & 0 & 0 & 0 & 0 & 0 \\ c_1^{-N_{\text{Air}}} & c_2^{-N_{\text{Air}}} & c_3^{-N_{\text{Air}}} & c_4^{-N_{\text{Air}}} & c_5^{-N_{\text{Air}}} & c_6^{-N_{\text{Air}}} & c_7^{-N_{\text{Air}}} & c_8^{-N_{\text{Air}}} & c_9^{-N_{\text{Air}}} \\ c_1^{-N_{\text{Air}}+1} & c_2^{-N_{\text{Air}}+1} & c_3^{-N_{\text{Air}}+1} & c_4^{-N_{\text{Air}}+1} & c_5^{-N_{\text{Air}}+1} & c_6^{-N_{\text{Air}}+1} & c_7^{-N_{\text{Air}}+1} & c_8^{-N_{\text{Air}}+1} & c_9^{-N_{\text{Air}}+1} \\ c_1^{0} & c_2^{0} & c_3^{0} & c_4^{0} & c_5^{0} & c_6^{0} & c_7^{0} & c_8^{0} & c_9^{0} \\ c_1^{N_{\text{Air}}-1} & c_2^{N_{\text{Air}}-1} & c_3^{N_{\text{Air}}-1} & c_4^{N_{\text{Air}}-1} & c_5^{N_{\text{Air}}-1} & c_6^{N_{\text{Air}}-1} & c_7^{N_{\text{Air}}-1} & c_8^{N_{\text{Air}}-1} & c_9^{N_{\text{Air}}-1} \\ c_1^{N_{\text{Air}}} & c_2^{N_{\text{Air}}} & c_3^{N_{\text{Air}}} & c_4^{N_{\text{Air}}} & c_5^{N_{\text{Air}}} & c_6^{N_{\text{Air}}} & c_7^{N_{\text{Air}}} & c_8^{N_{\text{Air}}} & c_9^{N_{\text{Air}}} \end{pmatrix}$$

**c**

$$\mathbf{A} = \mathcal{M}_{\text{sub}}^{T_1 \oplus U} \oplus \mathcal{M}_{\text{EO}}^{T_2} + CK_P \qquad \phi_i = C \sum_j (\mathbf{P}_{\text{Air}} \mathbf{V}_{\text{sub}})_{i,j} = C \sum_j^U \mathbf{V}_{\text{sub}; ij} \qquad \Phi_{i,j} = \phi_i$$

$$\mathbf{A} = \begin{pmatrix} \mathcal{M}_{\text{sub}}^{T_1 \oplus U} & \mathbb{0} \\ \mathbb{0} & \mathcal{M}_{\text{EO}}^{T_2} \end{pmatrix} + \begin{pmatrix} \mathbb{0} & \Phi \\ \Phi^{\text{Tr}} & \mathbb{0} \end{pmatrix}$$

$$\mathbf{A} = \begin{pmatrix} \mathcal{M}_{\text{sub}}^{-N_{\text{sub}}} & 0 & 0 & 0 & 0 & 0 & 0 & 0 & 0 & \varphi_{-N_s} & \varphi_{-N_s} & \varphi_{-N_s} & \varphi_{-N_s} \\ 0 & \mathcal{M}_{\text{sub}}^{-N_{\text{sub}}+1} & 0 & 0 & 0 & 0 & 0 & 0 & 0 & \varphi_{-N_s+1} & \varphi_{-N_s+1} & \varphi_{-N_s+1} & \varphi_{-N_s+1} \\ 0 & 0 & \mathcal{M}_{\text{sub}}^{-N_{\text{sub}}+2} & 0 & 0 & 0 & 0 & 0 & 0 & \varphi_{-N_s+2} & \varphi_{-N_s+2} & \varphi_{-N_s+2} & \varphi_{-N_s+2} \\ 0 & 0 & 0 & \mathcal{M}_{\text{sub}}^{-N_{\text{sub}}+3} & 0 & 0 & 0 & 0 & 0 & \varphi_{-N_s+3} & \varphi_{-N_s+3} & \varphi_{-N_s+3} & \varphi_{-N_s+3} \\ 0 & 0 & 0 & 0 & \mathcal{M}_{\text{sub}}^{0} & 0 & 0 & 0 & 0 & \varphi_{0} & \varphi_{0} & \varphi_{0} & \varphi_{0} \\ 0 & 0 & 0 & 0 & 0 & \mathcal{M}_{\text{sub}}^{N_{\text{sub}}-3} & 0 & 0 & 0 & \varphi_{N_s-3} & \varphi_{N_s-3} & \varphi_{N_s-3} & \varphi_{N_s-3} \\ 0 & 0 & 0 & 0 & 0 & 0 & \mathcal{M}_{\text{sub}}^{N_{\text{sub}}-2} & 0 & 0 & \varphi_{N_s-2} & \varphi_{N_s-2} & \varphi_{N_s-2} & \varphi_{N_s-2} \\ 0 & 0 & 0 & 0 & 0 & 0 & 0 & \mathcal{M}_{\text{sub}}^{N_{\text{sub}}-1} & 0 & \varphi_{N_s-1} & \varphi_{N_s-1} & \varphi_{N_s-1} & \varphi_{N_s-1} \\ 0 & 0 & 0 & 0 & 0 & 0 & 0 & 0 & \mathcal{M}_{\text{sub}}^{N_{\text{sub}}} & \varphi_{N_s} & \varphi_{N_s} & \varphi_{N_s} & \varphi_{N_s} \\ \varphi_{-N_s} & \varphi_{-N_s+1} & \varphi_{-N_s+2} & \varphi_{-N_s+3} & \varphi_0 & \varphi_{N_s-3} & \varphi_{N_s-2} & \varphi_{N_s-1} & \varphi_{N_s} & \mathcal{M}_{\text{EO}}^{-N_{\text{EO}}} & 0 & 0 & 0 \\ \varphi_{-N_s} & \varphi_{-N_s+1} & \varphi_{-N_s+2} & \varphi_{-N_s+3} & \varphi_0 & \varphi_{N_s-3} & \varphi_{N_s-2} & \varphi_{N_s-1} & \varphi_{N_s} & 0 & \mathcal{M}_{\text{EO}}^{-N_{\text{EO}}+1} & 0 & 0 \\ \varphi_{-N_s} & \varphi_{-N_s+1} & \varphi_{-N_s+2} & \varphi_{-N_s+3} & \varphi_0 & \varphi_{N_s-3} & \varphi_{N_s-2} & \varphi_{N_s-1} & \varphi_{N_s} & 0 & 0 & \mathcal{M}_{\text{EO}}^{N_{\text{EO}}-1} & 0 \\ \varphi_{-N_s} & \varphi_{-N_s+1} & \varphi_{-N_s+2} & \varphi_{-N_s+3} & \varphi_0 & \varphi_{N_s-3} & \varphi_{N_s-2} & \varphi_{N_s-1} & \varphi_{N_s} & 0 & 0 & 0 & \mathcal{M}_{\text{EO}}^{N_{\text{EO}}} \end{pmatrix}$$

**d**

$$\mathbf{V}^{-1} \mathbf{A} \mathbf{V} = \mathcal{M}$$

$$\mathbf{P}_{\text{EO}} = \mathbb{1}^{T_1} \oplus \mathbb{0}^U \oplus \mathbb{1}^{T_2} = \begin{pmatrix} \mathbb{1} & \mathbb{0} & \mathbb{0} \\ \mathbb{0} & \mathbb{0} & \mathbb{0} \\ \mathbb{0} & \mathbb{0} & \mathbb{1} \end{pmatrix} \qquad \mathbf{P}_{\text{Air}} = \mathbb{0}^{T_1} \oplus \mathbb{1}^U \oplus \mathbb{0}^{T_2} = \begin{pmatrix} \mathbb{0} & \mathbb{0} & \mathbb{0} \\ \mathbb{0} & \mathbb{1} & \mathbb{0} \\ \mathbb{0} & \mathbb{0} & \mathbb{0} \end{pmatrix}$$

$$\chi^2_{\text{EO};i} = \sum_j (\mathbf{P}_{\text{EO}} \mathbf{V} \circ \mathbf{V})_{i,j} = \sum_j^{T_1, T_2} V_{i,j}^2 \qquad \chi^2_{\text{Air};i} = \sum_j (\mathbf{P}_{\text{Air}} \mathbf{V} \circ \mathbf{V})_{i,j} = \sum_j^U V_{i,j}^2$$

**Extended Data Fig. 6| Coupled-Oscillator Model – Coupling Matrix Formalism. a,** Depiction of the coupling matrix for the sub-cavity, coupling one EO crystal with the air gap. It is represented here using $N_{\text{Air}} = N_{\text{EOC}} = 2$, resulting in a matrix dimension of $N_{\text{sub}} = (2N_{\text{Air}} + 1) + 2N_{\text{EO}} = 9$, partitioned into the blocks filled by the original resonances, and the coupling sub-blocks. **b,** The eigenvectors from the sub-cavity coupling matrix are depicted, as well as the result from projection onto the air sub-space, as is needed to construct the coupling matrix for the full cavity, as shown in panel c **c,** The full coupling matrix is depicted, which is built from the sub-blocks of the final EO crystal and the resultant eigenvalues from the sub-cavity, where these two blocks are then coupled via the same coupling matrix as in the sub-cavity, excepted for a modification by the integrated projection of the sub-cavity eigenvectors onto the air sub-space. We term the matrix collecting these integrated projections times the coupling constant $\Phi$, where for the i[th] eigenvector of the sub-cavity system, its integrated projection onto the air sub-space is then $\phi_i/C$. **d,** The character of the final resulting eigenvectors is calculated either by using the projection operator onto any particular sub-space on the matrix consisting the square of the eigenvectors (i.e. the Hadamard product of $\mathbf{V}$ with itself), or equivalently by performing the summation over the squared elements of the matrix $\mathbf{V}$ within the appropriate sub-space.

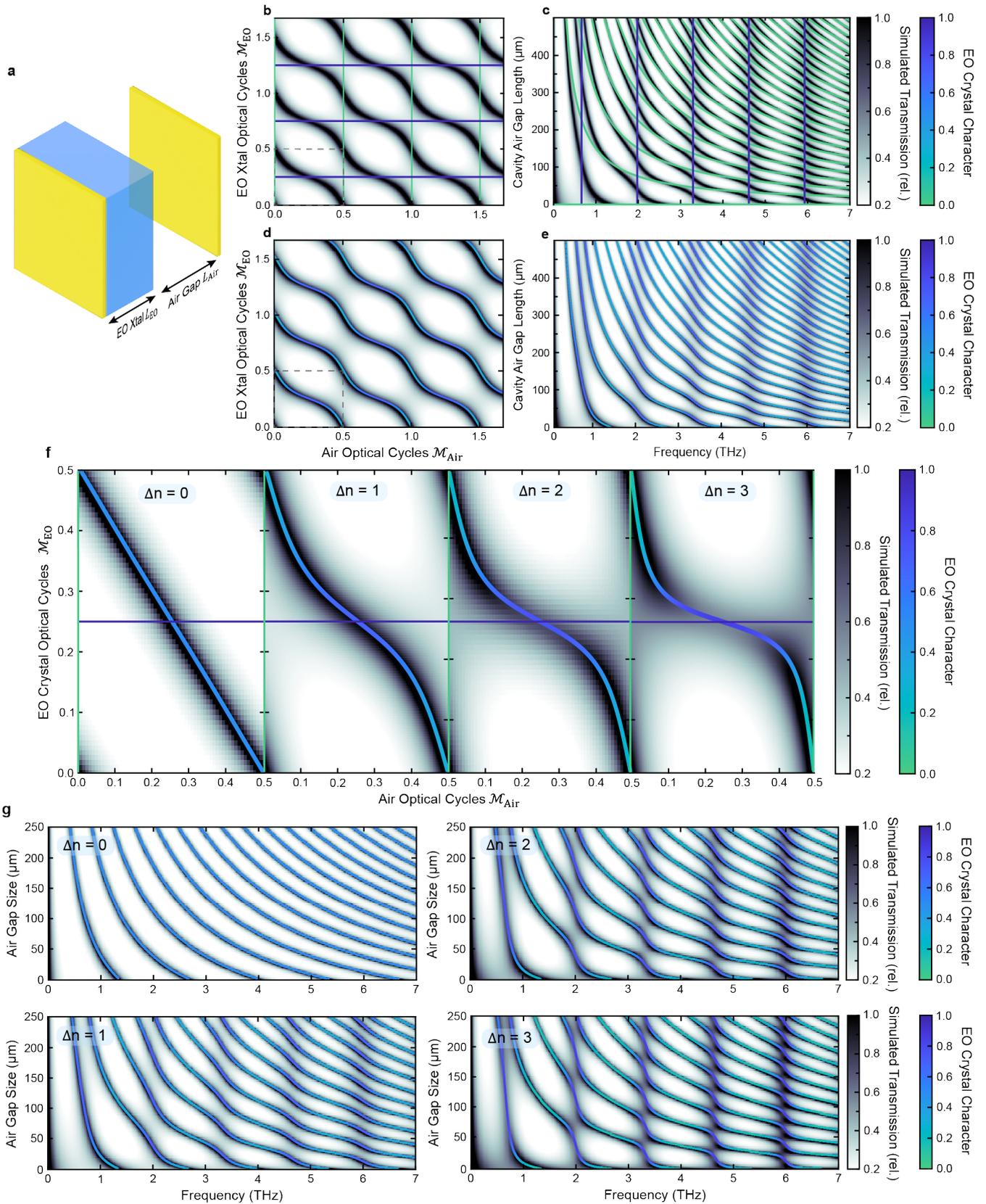

**Extended Data Fig. 7| Coupled-Oscillator Model – Partial Cavity. a,** Depiction of the partial cavity analyzed here, where there is a single EO crystal of thickness $L_{EO}$ and two gold mirrors, separated by a variable-size air gap $L_{Air}$. **b,** Simulated transmission as a function of number of optical cycles in air and the EO crystal, for $n_{EO} = 2$, along with the basis resonances, colored according to the EO character. The irreducible space is denoted by the dashed gray square. **c,** Simulated transmission as a function of number of air gap size and THz frequency, for $n_{EO} = 2$, along with the corresponding basis resonances, colored according to the EO character. **d,** The eigenvalues are plotted on top of the simulated transmission displayed in panel b, colored here according to the resultant EO character. **e,** The eigenvalues are plotted on top of the simulated transmission displayed in panel c, colored according to the resultant EO character. **f,** The eigenvalues are plotted in the irreducible space, as a function of EO crystal refractive index, referenced here as a difference according to the air refractive index of unity, highlighting maximal coupling at no index mismatch, and a shrinking coupling magnitude as the refractive index mismatch becomes larger. **g,** A representative cavity dispersion is plotted for each sub-panel in panel f, where the optical thickness has been normalized, to highlight the behavior of the mode coupling strength as a function of varying refractive index. In each sub-panel, the simulated cavity transmission is displayed in gray, and the eigenvalues obtained from the coupled-oscillator model are displayed, and colored according to the EO crystal character.

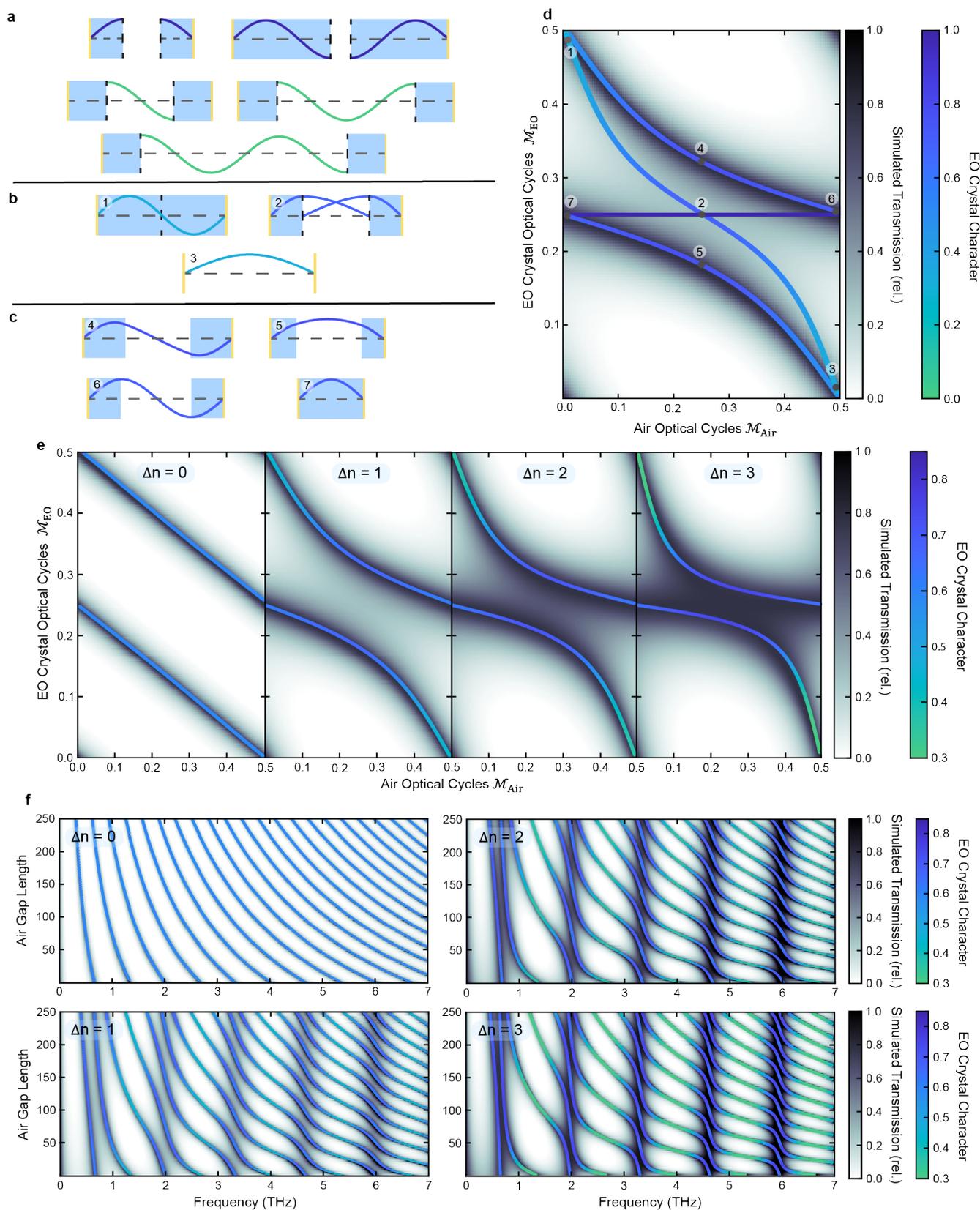

**Extended Data Fig. 8| Coupled-Oscillator Model – Full Cavity. a,** $\lambda/4$ standing-wave resonances in EO crystals (purple) and $\lambda/2$ standing-wave resonances in the air gap (green), all for a common frequency, with variable EO crystal lengths (top) and air gap sizes (bottom). **b,** Selected resonance conditions derived from the sub-cavity eigenvectors, colored according to the EO character (scale in panel d), and depicted at illustrative cavity conditions, corresponding to labels in panel d. **c,** Resonance conditions corresponding to the full-cavity eigenvectors, colored according to the derived EO character (scale in panel d), depicted at illustrative cavity conditions labelled in panel d. **d,** Simulated field transmission (false color, gray), along with resonances corresponding to the sub-cavity (interior eigenfunction), and the symmetric (below) and anti-symmetric (above) eigenvectors, colored according to the corresponding quartz character. **e,** The simulated transmission for a series of EO crystal refractive indices, referenced according to refractive index of air ($n_{Air} = 1$), to demonstrate the relative 'de-coupling' of the air and EO-crystal modes for high refractive indices. **f,** The frequency dispersion of the optical modes is shown for the refractive indices corresponding to panel e, representing the evolution of the coupling strength as it would be observed experimentally.

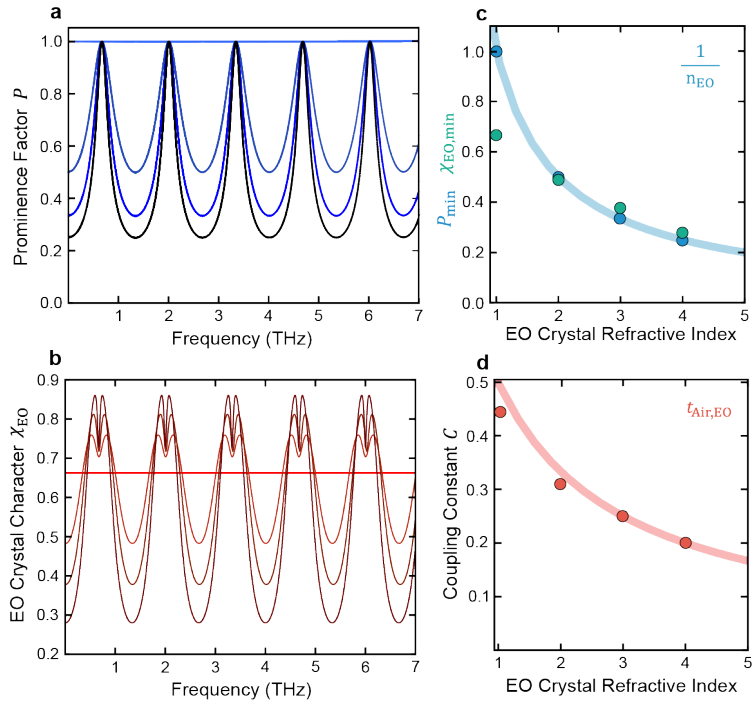

**Extended Data Fig. 9| Effect of Electro-Optic Crystal Refractive Index**. a, The prominence factor is depicted, for the electro-optic crystal refractive indices of 1,2,3,4 with decreasing saturation. **b,** The electro-optic crystal character identified from the coupled-oscillator model are plotted for the same input refractive indices as considered in panel a. **c,** The coupling constant identified from the coupled-oscillator model is plotted for the various EO crystal refractive indices that were investigated, and compared with the transmission coefficient from the air gap into the EO crystal, displaying very good correspondence. **d,** The minimum value observed in both the prominence factor and EO crystal character are displayed as a function of EO crystal refractive index, where we note that the computed prominence factor minimum is exactly equal to the inverse of the refractive index, whereas the EO crystal character follows a very similar trend.
20

**Table 1| Quartz Dielectric Function.** Values adapted and expanded from Frenzel, et.al.

$$\varepsilon(f) = 2.103 + \sum_l \frac{A_l f_l^2}{f_l^2 - f^2 - i\Gamma_l f}$$

| Mode Index | Central Frequency $f_l$ (THz) | Amplitude $A_l$ | Damping Factor $\Gamma_l$ (THz) |
|---|---|---|---|
| Oscillator 1 | 3.85 | 0.000055 | 0.09 |
| Oscillator 2 | 7.97 | 0.032 | 0.167 |
| Oscillator 3 | 11.76 | 0.37 | 0.096 |
| Oscillator 4 | 13.45 | 0.75 | 0.177 |
| Oscillator 5 | 24 | 0.14 | 0.6 |
| Oscillator 6 | 37.2 | 0.6 | 0.6 |

**Table 2| Quartz 2nd Order Nonlinear Susceptibility.** Values adapted from Frenzel, et.al., where available, and otherwise inferred via standard electro-optic measurements.

$$\chi_{eff}^{(2)}(f) = 0.28 \cdot \left(1 + \sum_l \frac{C_l f_l^2}{f_l^2 - f^2 - i\Gamma_l f}\right) \frac{\text{pm}}{\text{V}}$$

| Mode Index | Central Frequency $f_l$ (THz) | Faust-Henry Coef. $C_l$ | Damping Factor $\Gamma_l$ |
|---|---|---|---|
| Oscillator 1 | 3.85 | 0.11* | 0.09 |
| Oscillator 2 | 7.97 | -0.012 | 0.167 |
| Oscillator 3 | 11.76 | 0.5 | 0.096 |
| Oscillator 4 | 13.45 | -0.6 | 0.177 |
| Oscillator 5 | 24 | 1.2 | 0.6 |

\* $C_1$ found to vary widely from crystal to crystal, by up to nearly 50%

# Supplementary Information:
# Electro-Optic Cavities for In-Situ Measurement of Cavity Fields

**Michael S. Spencer**[1*], **Joanna M. Urban**[1], **Maximilian Frenzel**[1], **Niclas S. Mueller**[1], **Olga Minakova**[1], **Martin Wolf**[1], **Alexander Paarmann**[1], **Sebastian F. Maehrlein**[1,2,3*]

1. Department of Physical Chemistry, Fritz Haber Institute of the Max Planck Society, 14195 Berlin, Germany
2. Helmholtz-Zentrum Dresden-Rossendorf, Institute of Radiation Physics, 01328 Dresden, Germany
3. Technische Universität Dresden, Institute of Applied Physics, 01062 Dresden, Germany

*email: maehrlein@fhi-berlin.mpg.de, spencer@fhi-berlin.mpg.de

## Table of Contents

## Supplementary Discussion 1
Optical Properties of Gold Films on z-cut α-Quartz

We investigate the dispersive optical properties of the gold films deposited onto the quartz crystals using a three-layer dielectric model. We consider the reflection or transmission of a pulse from the two interfaces, i.e. the air-gold and gold-quartz interfaces (see Supplementary Fig. 1 below), calculated at normal incidence. Only the principal pulse is considered, i.e. not the subsequent cavity reflections within quartz, as these are discussed subsequently in Supplementary Discussion 2. The separation of these two optical processes is justified because the gold film is deeply sub-wavelength at the thicknesses considered here.

In the following analysis, we have suppressed the variable with respect to frequency, for brevity, but all transmission and reflections are inherently a function of frequency due to the dispersive refractive indices of quartz and gold. We now report all possible terms that exist in the transmission and reflection regions, considering a pulse propagating starting from either outside (external $r_\text{ext}$, $t_\text{ext}$, panel a) or inside (internal: $r_\text{int}$, $t_\text{int}$, panel b) the cavity:

$$r_\text{ext} \equiv \frac{E_\text{ref}}{E_\text{inc}^\text{ext}} = r_\text{Air,Au} + t_\text{Au,Air}r_\text{Au,Qtz}t_\text{Air,Au}e^{2i\phi} + t_\text{Au,Air}r_\text{Au,Air}r_\text{Au,Qtz}^2 t_\text{Air,Au}e^{4i\phi} + t_\text{Au,Air}r_\text{Au,Air}^2 r_\text{Au,Qtz}^3 t_\text{Air,Au}e^{6i\phi} + \cdots, \quad (S1.1)$$

$$t_\text{ext} \equiv \frac{E_\text{trans}}{E_\text{inc}^\text{ext}} = t_\text{Au,Qtz}t_\text{Air,Au}e^{i\phi} + t_\text{Au,Qtz}r_\text{Au,Air}r_\text{Au,Qtz}t_\text{Air,Au}e^{3i\phi} + t_\text{Au,Qtz}r_\text{Au,Air}^2 r_\text{Au,Qtz}^2 t_\text{Air,Au}e^{5i\phi} + \cdots, \quad (S1.2)$$

$$r_\text{int} = \frac{E_\text{ref}}{E_\text{inc}^\text{int}} \equiv r_\text{Qtz,Au} + t_\text{Au,Qtz}r_\text{Au,Air}t_\text{Qtz,Au}e^{2i\phi} + t_\text{Au,Qtz}r_\text{Au,Qtz}r_\text{Au,Air}^2 t_\text{Qtz,Au}e^{4i\phi} + t_\text{Au,Air}r_\text{Au,Air}^2 r_\text{Au,Qtz}^3 t_\text{Air,Au}e^{6i\phi} + \cdots, \quad (S1.3)$$

$$t_\text{int} = \frac{E_\text{trans}}{E_\text{inc}^\text{int}} \equiv t_\text{Au,Air}t_\text{Qtz,Au}e^{i\phi} + t_\text{Au,Air}r_\text{Au,Qtz}r_\text{Au,Air}t_\text{Qtz,Au}e^{3i\phi} + t_\text{Au,Air}r_\text{Au,Qtz}^2 r_\text{Au,Air}^2 t_\text{Qtz,Au}e^{5i\phi} + \cdots. \quad (S1.4)$$

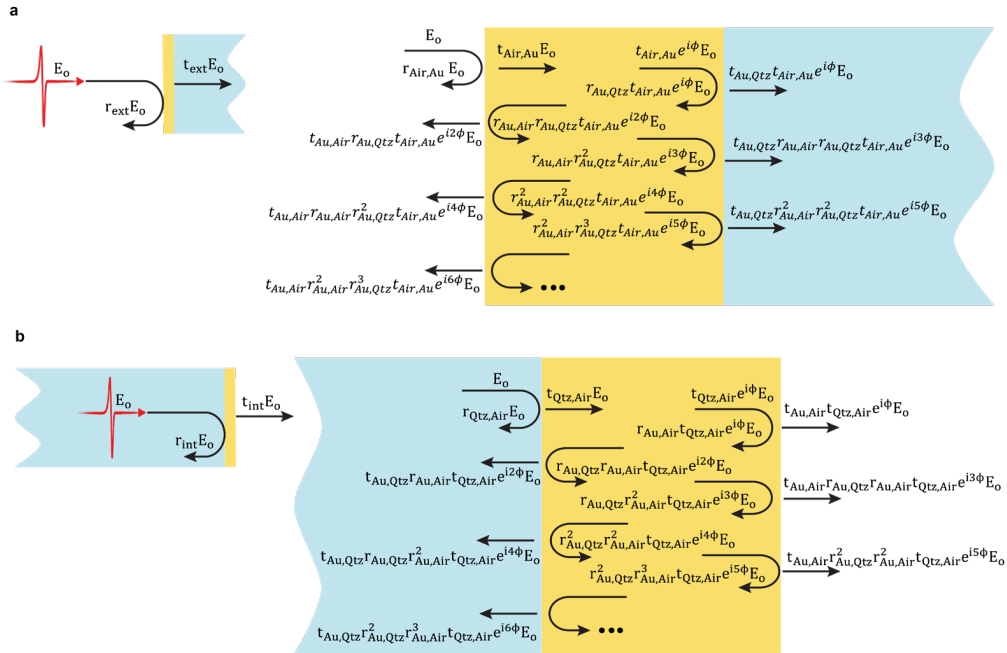

**Supplementary Fig. 1| Internal and External Reflections From Cavity Mirrors. a,** The external reflection and transmission of a pulse (left) is diagrammed (right) as resulting from all possible internal reflections and transmissions from the thin gold layer. The summation of all terms propagating left constitutes the external reflection coefficient (S1.1), and all rightward propagating constitutes the external transmission coefficient (S1.2). **b,** The internal reflection and transmission of a pulse (left) is diagrammed (right) as resulting from all possible internal reflections and transmissions from the thin gold layer. The summation of all terms propagating left constitutes the internal reflection coefficient (S1.3), and all rightward propagating constitutes the external transmission coefficient (S1.4).

We define the (frequency-dependent) transmission or reflection coefficients in terms of the ratio of the spectrum of the transmitted or reflected pulse with the spectrum of the relevant incident pulse. The reflection and transmission coefficient at a given interface is therefore given by $r_{i,j} = (n_i - n_j)/(n_i + n_j)$ and $t_{i,j} = 2n_i/(n_i + n_j)$, respectively, and the phase accumulation and attenuation over the gold film length is defined as $\phi = k(\omega)d_\text{Au} = \omega(n_\text{Au}(\omega) + i\kappa_\text{Au}(\omega))d_\text{Au}/c_\text{o}$, where $c_\text{o}$ is the vacuum speed of light, $n$ and $\kappa$ are the real and imaginary components of the refractive index, and where we have used the plane-wave convention $E_\text{o}e^{i(kz-\omega t)}$ for a rightward-traveling ($+z$) plane wave. Summation using the geometric series allows the above expressions to be evaluated as the following:

$$r_\text{ext} \equiv \frac{E_\text{ref}}{E_\text{inc}^\text{ext}} = r_\text{Air,Au} + \frac{t_\text{Au,Air}r_\text{Au,Qtz}t_\text{Air,Au}e^{2i\phi}}{1 - r_\text{Au,Air}r_\text{Au,Qtz}e^{2i\phi}}, \quad (S1.5)$$

$$t_{\text{ext}} \equiv \frac{E_{\text{trans}}}{E_{\text{inc}}^{\text{ext}}} = \frac{t_{\text{Au,Qtz}}t_{\text{Air,Au}}e^{i\phi}}{1 - r_{\text{Au,Air}}r_{\text{Au,Qtz}}e^{2i\phi}}, \tag{S1.6}$$

$$r_{\text{int}} = \frac{E_{\text{ref}}}{E_{\text{inc}}^{\text{int}}} = r_{\text{Qtz,Au}} + \frac{t_{\text{Au,Qtz}}r_{\text{Au,Air}}t_{\text{Qtz,Au}}e^{2i\phi}}{1 - r_{\text{Au,Qtz}}r_{\text{Au,Air}}e^{2i\phi}}, \tag{S1.7}$$

$$t_{\text{int}} = \frac{E_{\text{trans}}}{E_{\text{inc}}^{\text{int}}} = \frac{t_{\text{Au,Air}}t_{\text{Qtz,Au}}e^{i\phi}}{1 - r_{\text{Au,Qtz}}r_{\text{Au,Air}}e^{2i\phi}}. \tag{S1.8}$$

These functions are plotted in Supplementary Figure 2, for various gold thickness, spanning the extremes of 0 to 200 nm, using the complex, frequency-dependent refractive indices computed from the quartz dielectric function used in Table 1 of Methods and the Drude Parameters for gold reported in the main text. We display the absolute values of the coefficients in Supplementary Fig. 2, but note that they are generally complex. We also note that there is significant dispersion in these transmission functions (both internal and external), which is a fundamental optical feature of the multi-layer structure, and could only be eliminated in the case of a true, free-standing gold film. Therefore, pulse dispersion should be considered in any case where the gold film is deposited on a dielectric substrate, and where detailed knowledge of the precise electric field (i.e. nonlinear field-driven effects) is of critical importance. In Extended Data Figure 3d,c, we demonstrate the consequences of this, showing the relative increase in pulse FWHM and central frequency for pulses transmitted into and out of the cavity, presenting a technical hurdle to connecting an externally-measured field with the internal one within a cavity. This effect persists until much thicker films, ($d_{\text{Au}} \sim$ 100 nm) at which point the optical path length in the gold films becomes comparable to the wavelength corresponding to frequencies of a few THz. In this limit, the absorption in gold becomes the dominant feature in the spectral properties of the multi-layer interface.

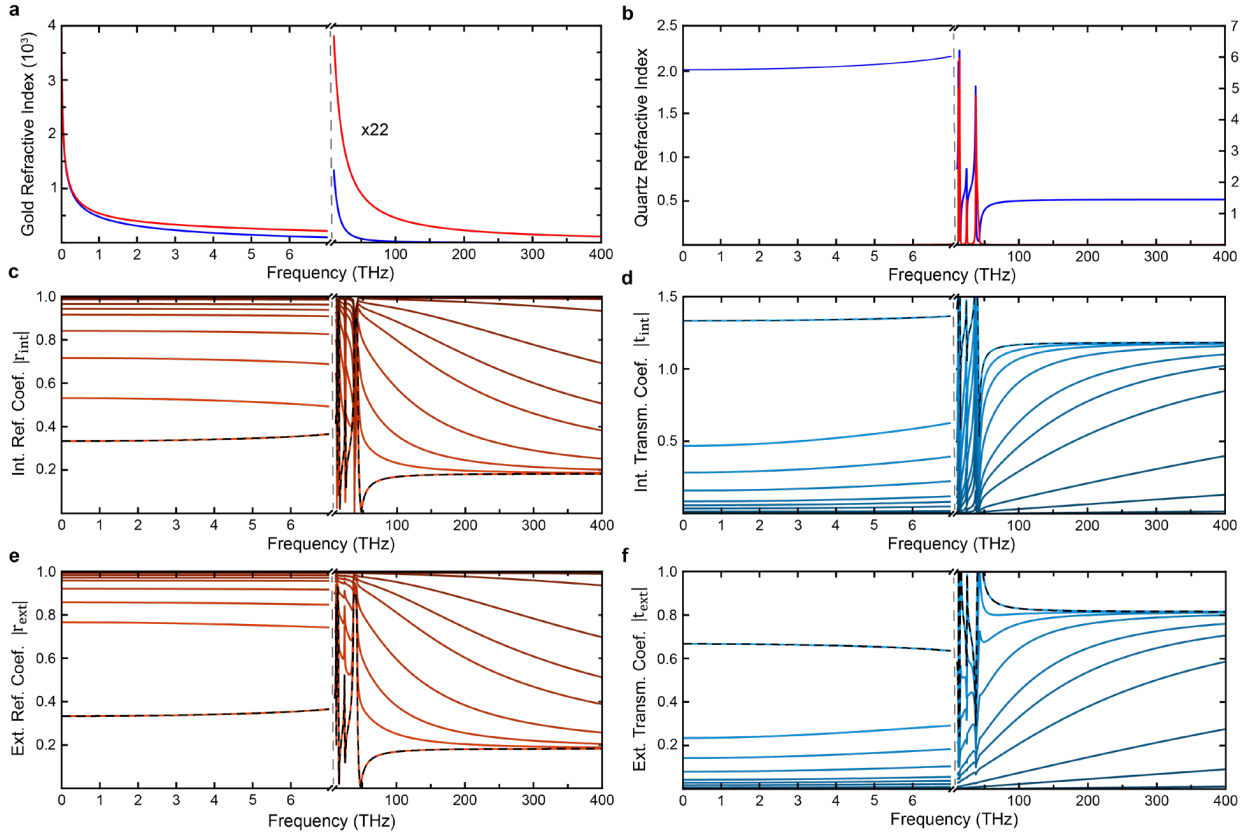

**Supplementary Fig. 2| Optical Properties of Cavity Mirrors. a**, Complex dielectric function of gold. The real part is displayed in blue, and the imaginary part in red. **b**, Complex refractive index function of quartz. The real part is displayed in blue, and the imaginary part in red. **c**, Absolute value of the internal field reflection coefficient (S1.7; absolute value) as a function of gold film thickness, plotted for the values 0, 0.5, 1, 2, 4, 6, 10, 25, 50, 100, and 200 nm, from bottom to top. The result for $n_{Qtz,Air}$ is plotted as well (dotted black line), in correspondence with the $d_{Au}$ = 0 nm result. **d**, Absolute value of the internal field transmission coefficient (S1.8; absolute value) is plotted for the same thickness parameters as in panel c. **e**, Absolute value of the external field reflection coefficient (S1.5; absolute value) is plotted for the same thickness parameters as in panel c. **f**, Absolute value of the external field transmission coefficient (S1.6; absolute value) is plotted for the same thickness parameters as in panel c.

## Supplementary Discussion 2
<u>Cavity Electric Fields: Travelling- and Standing-Wave Pictures</u>

One way to describing an electric field inside of a cavity is the picture of a traveling pulse, experiencing successive reflections at every interface. The complementary picture to the traveling pulse is that of the 'cavity modes', obtained via Fourier transformation, which can function as an eigenbasis of the cavity electric field. The equivalence of these pictures is demonstrated in this section. We write the incident THz pulse, $E_{\text{THz}}^{\text{inc}}(t,z)$, - here, in terms of its Fourier-transform $E_{\text{THz}}^{\text{inc}}(\omega, k_z)$ - as a simplified, one-dimensional function of time (frequency) and space (momentum), where the inclusion of the momentum, $k_z(\omega) = \omega n(\omega)/c_o$, is necessary to include effects of dispersion on the pulse inside the cavity medium:

$$E_{\text{THz}}^{\text{inc}}(t,z) = \frac{1}{\sqrt{2\pi}} \int E_{\text{THz}}^{\text{inc}}(\omega, k_z)\, e^{i(k_z(\omega)z - \omega t)}\, d\omega \,. \tag{S2.1}$$

Supplementary Figure S3 demonstrates how a wave traveling within the cavity will progressively decrease in amplitude, due to the partial transmission of the field at the imperfect gold mirrors. Each partial reflection amplitude is related to the original incident wave, using both the internal and external reflection and transmission field coefficients, which are detailed in Supplementary Discussion 1.

To compare with electro-optic sampling measurements, we want to write down the cavity field as a function of both space and frequency. To do this, we write down the electric field at an arbitrary cavity position, and detail the form arising due to the cavity end-mirror reflections, as a function of time. We do this, while including the effects of dispersion, using the Fourier transform definition (Equation S2.1):

$$E_{\text{THz}}^{\text{cav}}(t,z) = t_{\text{ext}} \left[ \sum_{q=0}^{\infty} r_{\text{int}}^{2q} E_{\text{THz}}^{\text{inc}}\left( t - q\tau_{\text{RT}}(\omega) + \frac{zn(\omega)}{c_o}, z \right) + \sum_{q=0}^{\infty} r_{\text{int}}^{2q+1} E_{\text{THz}}^{\text{inc}}\left( t - \frac{(2q+1)\tau_{\text{RT}}(\omega)}{2} - \frac{zn(\omega)}{c_o}, z \right) \right]. \tag{S2.2}$$

where the first summation is all forward-propagating instances of the pulse passing at some position $z$ in the cavity, and the second summation is the backward-propagating instances at that same position, and where $\tau_{\text{RT}}(\omega) = 2L_{\text{cav}} n(\omega)/c_o$ is the cavity round-trip time. By Fourier transforming this expression, we write the spatially-resolved cavity spectrum as it relates to the incident pulse spectrum:

$$E_{\text{THz}}^{\text{inc}}(\omega, z) = E_{\text{THz}}^{\text{inc}}(\omega) t_{\text{ext}} \left[ \sum_{q=0}^{\infty} r_{\text{int}}^{2q}\, e^{i(k_z(\omega)z - q\omega\tau_{\text{RT}})} + \sum_{q=0}^{\infty} r_{\text{int}}^{2q+1}\, e^{i\left(-k_z(\omega)z - \left(\frac{2q+1}{2}\right)\omega\tau_{\text{RT}}\right)} \right]. \tag{S2.3}$$

The infinite summations can be carried out analytically using the geometric series definition(s), yielding:

$$\frac{E_{\text{THz}}^{\text{cav}}(\omega, z)}{E_{\text{THz}}^{\text{inc}}(\omega, z)} = t_{\text{ext}} \left[ \frac{e^{ik_z(\omega)z}}{1 - r_{\text{int}}^2 e^{-i\omega\tau_{\text{RT}}(\omega)}} + \frac{r_{\text{int}} e^{-i\left(k_z(\omega)z + \frac{\omega\tau_{\text{RT}}(\omega)}{2}\right)}}{1 - r_{\text{int}}^2 e^{-i\omega\tau_{\text{RT}}(\omega)}} \right] = \frac{t_{\text{ext}}\left( e^{ik_z(\omega)z} + r_{\text{int}} e^{-ik_z(\omega)z} e^{\frac{-i\omega\tau_{\text{RT}}(\omega)}{2}} \right)}{1 - r_{\text{int}}^2 e^{-i\omega\tau_{\text{RT}}(\omega)}}. \tag{S2.4}$$

Supplementary Figure 3 displays the time-domain fields, and corresponding amplitude- and phase-resolved spectrum of the fields evaluated at the cavity center ($z = 0$), comparing the total field to the purely forward- or backwards-propagating pulse summations, for the choices of $d_{\text{Au}} = 2$ nm, and $d_{\text{Au}} = 10$, all evaluated for a round-trip time $\tau_{\text{RT}} \approx 0.6$ ps, corresponding to the $L_{\text{Qtz}} = 44$ μm cavity. Note that for the simple derivation outlined above, we have suppressed any frequency content in the reflection and transmission coefficients for simplicity, but we use the full, dispersive functions outlined in Supplementary Discussion 1 when evaluating equation S2.4 to depict the fields in Supplementary Fig. 3. It is evident that constructive interference among the single-pulse Fourier transforms occurs only at the cavity resonance conditions, i.e. the frequencies which are integer multiples of the inverse of the round-trip time. In addition, we can see that only half of the modes are visible at the cavity position $z = 0$, due to perfect cancellation of Fourier transforms in the forward- and backward-propagating directions.

We now seek to show that this cavity response derived above can also be understood equivalently as a sum of cavity eigenmodes. To start, we show the cavity intensity response, considering either the total cavity field, or the forward/reverse-propagating cavity intensity responses:

$$\left| \frac{E_{\text{THz}}^{\text{cav}}(\omega, z)}{E_{\text{THz}}^{\text{inc}}(\omega, z)} \right|^2 = \frac{|t_{\text{ext}}|^2 \left( 1 + 2|r_{\text{int}}| \cos\left( k(\omega)z + \frac{\omega\tau_{\text{RT}}(\omega)}{2} \right) + |r_{\text{int}}|^2 \right)}{1 - 2|r_{\text{int}}|^2 \cos(\omega\tau_{\text{RT}}) + |r_{\text{int}}|^4}, \tag{S2.5}$$

$$\left| \frac{E_{\text{THz}}^{\text{cav,f}}(\omega, z)}{E_{\text{THz}}^{\text{inc}}(\omega, z)} \right|^2 = \frac{|t_{\text{ext}}|^2}{1 - 2|r_{\text{int}}|^2 \cos(\omega\tau_{\text{RT}}(\omega)) + |r_{\text{int}}|^4}, \quad \left| \frac{E_{\text{THz}}^{\text{cav,b}}(\omega, z)}{E_{\text{THz}}^{\text{inc}}(\omega, z)} \right|^2 = \frac{|t_{\text{ext}}|^2 |r_{\text{int}}|^2}{1 - 2|r_{\text{int}}|^2 \cos(\omega\tau_{\text{RT}}(\omega)) + |r_{\text{int}}|^4}. \tag{S2.6}$$

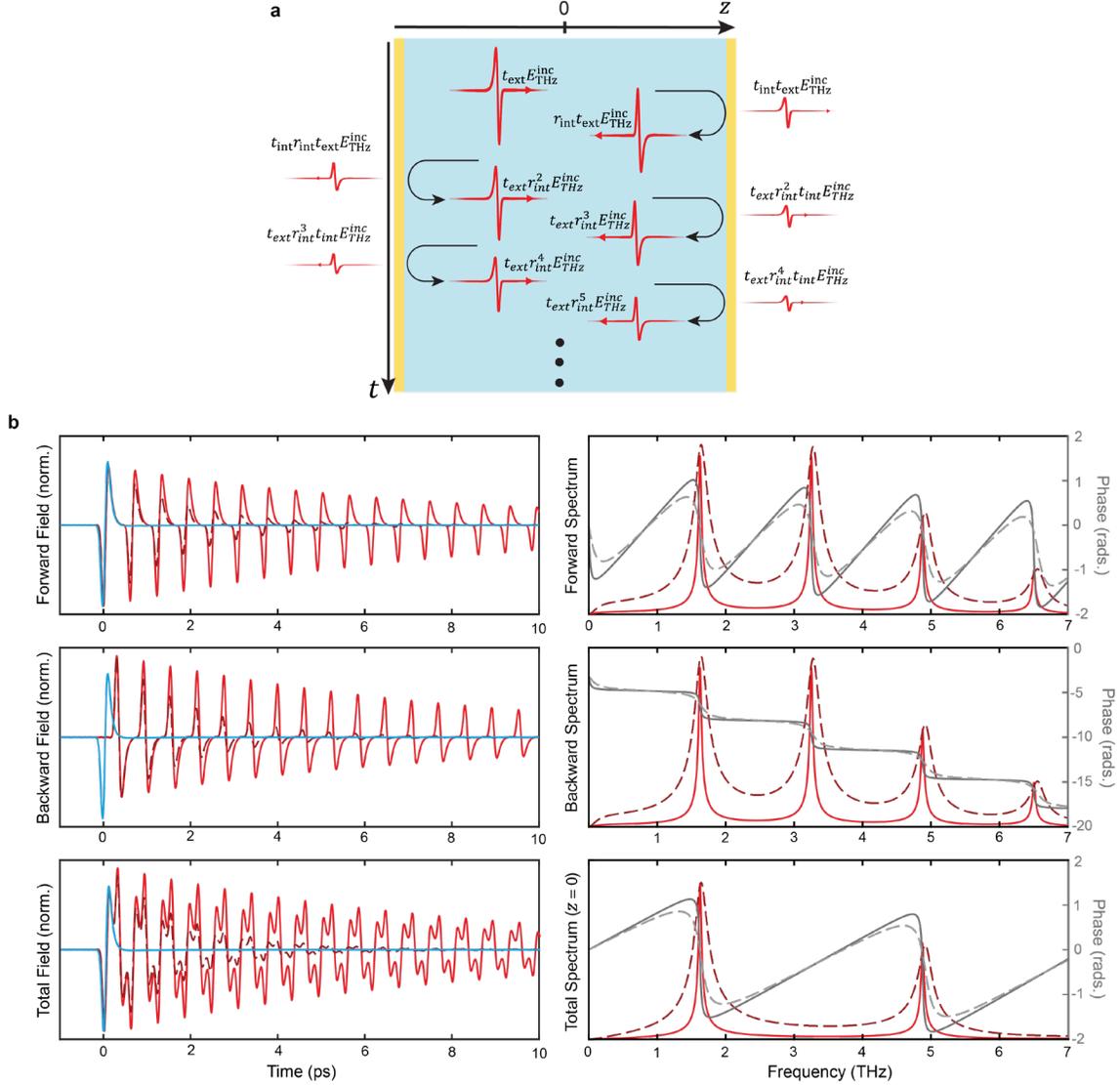

**Supplementary Fig. 3| Cavity Fields - Theory. a**, The internal cavity field is defined by the principal THz pulse reflecting internally within the cavity, partially transmitting at every encounter with the end mirrors. **b**, We plot the normalized, time-domain cavity electric fields (left), distinguishing between the forward-propagating instances (top) of the principal pulse (blue), the backwards-propagating instances (center), and their sum (bottom), in every case for gold film thicknesses of 2 (dashed lines) and 10 nm (solid lines). The corresponding spectra are shown on the right, represented in terms of the absolute value (red), and the phase (gray).

Moving forward, we consider only the forward-propagating field, for the sake of simplicity. Using the definition of the round-trip time, we can re-cast the cavity intensity response, using the speed of light $c = c_o/n(\omega)$, as the following:

$$\left|\frac{E_{\text{THz}}^{\text{cav,f}}(\omega,z)}{E_{\text{THz}}^{\text{inc}}(\omega,z)}\right|^2 = \frac{|t_{\text{ext}}|^2}{(1-|r_{\text{int}}|^2)^2 + 4|r_{\text{int}}|^2 \sin^2\left(\frac{\omega L_{\text{cav}}}{c}\right)} \quad . \tag{S2.7}$$

Finally, we now show that the sum of all cavity modes yields an identical expression to the one shown above, following the analysis of Ismail *et al.*[1]. To start, we note that all modes share the same loss rate as that of the travelling pulse, which is defined by the round-trip field loss rate and the cavity round trip time, as follows:

$$|r_{\text{int}}|^2 = e^{-\frac{\tau_{\text{RT}}}{\tau_{\text{loss}}}} \quad \rightarrow \quad \frac{1}{\tau_{\text{loss}}} = -\frac{\log(|r_{\text{int}}|^2)}{\tau_{\text{RT}}} \quad . \tag{S2.8}$$

The cavity modes have frequencies which correspond to integer multiples of the inverse of the cavity round trip time, i.e. $f_{\text{cav}}^q = \pm \frac{qc_o}{n(f_{\text{cav}}^q)L_{\text{Cav}}}$, where q is the mode index, which spans all integers. Using the loss rate relation, the electric field of these modes can be written as:

$$E_{\pm q}(t) = E_{\pm q}(0)\, e^{\mp 2\pi i f_{\text{cav}}^q t}\, e^{-\frac{t}{\tau_{\text{loss}}}}. \tag{S2.9}$$

From this, we can compute the Fourier transform of the modes, which has the typical Lorentzian form in terms of the field:

$$E_{\pm q}(\omega) = \frac{E_{\pm q}(0)}{\sqrt{2\pi}} \int_0^\infty e^{-i\omega_{\text{cav}}^q t} e^{\frac{-t}{\tau_{\text{loss}}}} e^{i\omega t} = \frac{E_{\pm q}(0)}{\sqrt{2\pi}} \left( \frac{1}{i(\omega \mp \omega_q) - \frac{1}{\tau_{\text{loss}}}} \right). \tag{S2.10}$$

From this, the cavity response for a particular mode can be calculated:

$$\left| \frac{E_{\pm q}(\omega)}{E_{\pm q}(0)} \right|^2 = \frac{1}{2\pi} \left( \frac{1}{(\omega \mp \omega_q)^2 - \frac{1}{\tau_{\text{loss}}^2}} \right). \tag{S2.11}$$

Normalizing against the linewidth, and introducing the term $\delta_{\text{out}} = -\log(r_{\text{int}}^2) = \tau_{\text{RT}}/\tau_{\text{loss}}$, representing the field reduction per round trip, we write the normalized cavity mode response as follows:

$$R_{\pm q} \equiv \frac{1}{\tau_{\text{loss}}} \left| \frac{E_{\pm q}(\omega)}{E_{\pm q}(0)} \right|^2 = \frac{1}{2\pi} \frac{\tau_{\text{RT}}}{\delta_{\text{out}}} \frac{\left(\frac{\delta_{\text{out}}}{\tau_{RT}}\right)^2}{(\omega \mp \omega_q)^2 + \left(\frac{\delta_{out}}{\tau_{RT}}\right)^2}. \tag{S2.12}$$

After several manipulations of the above expression[1], it can be shown that:

$$\left| \frac{E_{\text{THz}}^{\text{cav,f}}(\omega, z)}{E_{\text{THz}}^{\text{inc}}(\omega, z)} \right|^2 = \sum_{q=-\infty}^\infty \frac{1}{\tau_{\text{RT}}} R_{\pm q} = \frac{|t_{\text{ext}}|^4}{4 r_{\text{int}}^2 \sin^2\left(\frac{\omega n(\omega) L_{\text{cav}}}{c_o}\right) + (1 - r_{\text{int}}^2)^2}. \tag{S2.13}$$

Note that this expression is identical to the one derived using the time-domain analysis of a travelling pulse. Therefore, it can be demonstrated that the summation of all optical modes within the cavity produces the Fabry-Perot response derived from the travelling pulse formalism, showing the complete equivalence of the two pictures. In general, these two perspectives are totally equivalent, although in more complex systems the Fourier-domain analysis is certainly more mathematically straightforward.

Finally, we observe that the summation of the forwards- and backwards-propagating waves' interference leads to the familiar standing modes in space. This is demonstrated by evaluating the frequency-domain cavity field (Equation S2.4) at the resonance condition(s). The momenta of the cavity modes are given by $k_z(\omega_q) = \pi q / L_{\text{cav}}$, with corresponding wavelengths of $\lambda_q = 2n L_{\text{cav}}/q$, where we approximate without dispersion, leading to cavity mode frequencies $\omega_q = q \frac{\pi c_o}{L_{\text{cav}} n}$. Using these expressions, the cavity field simplifies to the following:

$$\frac{E_{\text{THz}}^{\text{cav}}(\omega, z)}{E_{\text{THz}}^{\text{inc}}(\omega)} = t_{\text{ext}} \frac{e^{i\left(\frac{\pi q}{L_{\text{cav}}}\right)z} + r_{\text{int}} e^{-i\left(\frac{\pi q}{L_{\text{cav}}}\right)z} e^{-i\pi q}}{1 - r_{\text{int}}^2}. \tag{S2.14}$$

We thus arrive at two different sets of solutions, assuming high internal reflectivites, depending on whether q is an even or odd integer:

$$E_{\text{THz}}^{\text{cav}}(\omega^{\text{even}}, z) \propto \cos\left(\frac{\pi q}{L_{\text{cav}}} z\right), \quad E_{\text{THz}}^{\text{cav}}(\omega^{\text{odd}}, z) \propto i \sin\left(\frac{\pi q}{L_{\text{cav}}} z\right) \tag{S2.15}$$

These standing waves, and the Lorentzian decomposition of the Fabry-Perot response are both demonstrated in Supplementary Figure S4. We note also the interesting property that in the case of no cavity, i.e. just an electro-optic crystal, that there is no loss in spectral amplitude for the cavity field at the constructive-interference conditional frequencies, if sampled at the correct cavity positions (see e.g. Supplementary Fig. 4b). Furthermore, we observe that once the cavity mirrors are included, the drop in peak spectral amplitude corresponds to $1/n_{\text{EO}}$.

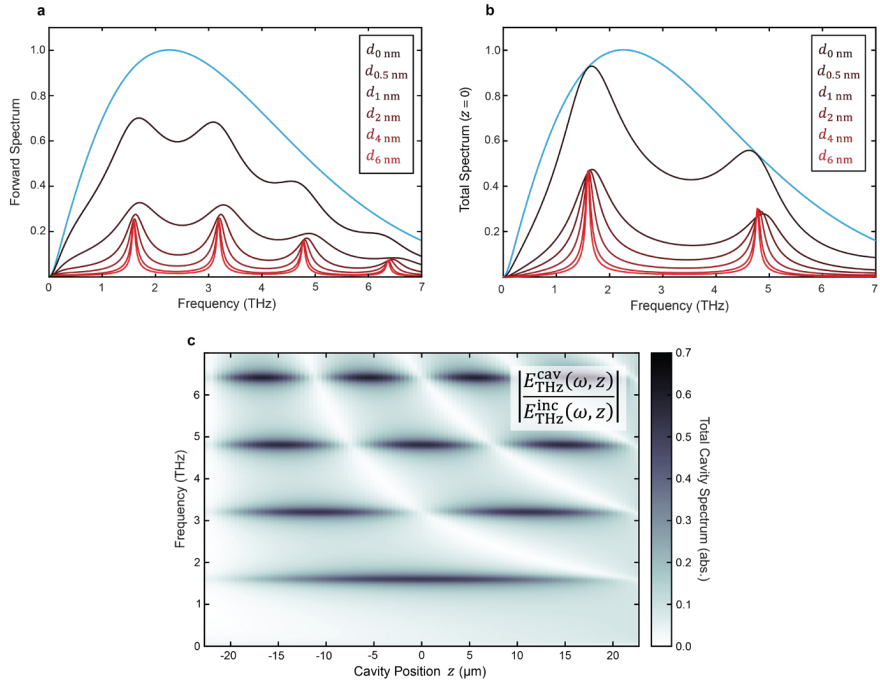

**Supplementary Fig. 4| Cavity Fields: Spatial Structure, Standing Wave. a**, The simulated cavity field spectra for strictly forward-propagation of the principal pulse are shown for various gold film thickness, all compared and normalized against the incident pulse spectrum (blue). **b**, The simulated cavity field spectra for the total cavity field is displayed, at the same gold thicknesses considered in panel a. **c,** The total cavity spectrum is shown as a function of cavity position and THz frequency. This simulation uses a gold film thickness of 2 nm, and a quartz crystal length of 44 µm.

## Supplementary Discussion 3
<u>Cavity Electro-Optic Sampling – Extended Discussion</u>

The cavity correction function utilized in the main text expands upon the conventional detector response function[2,3], which follows from considering a generic nonlinear wave mixing between terahertz and visible-frequency pulses, incorporating nonlinear boundary conditions and absorptive losses.[4,5] The general form of the cavity correction function is therefore understood as an extension of a conventional detector response function, except where we use the cavity electric fields of the THz and visible probing pulses to compute the nonlinear SFG/DFG polarizations in the cavity:

$$E_{\text{cav}}(\omega, z) = E_{\text{inc}}(\omega, z) \frac{t_{\text{ext}}(\omega) \left( e^{ik(\omega)z} + r_{\text{int}}(\omega) e^{-ik_z(\omega)z} e^{-\frac{i\omega \tau_{\text{RT}}(\omega)}{2}} \right)}{1 - r_{\text{int}}(\omega)^2 e^{-i\omega \tau_{RT}(\omega)}}. \tag{S3.1}$$

where here $\omega$ can refer to either the THz frequency $\Omega_{\text{THz}}$ or the visible frequency $\omega_{\text{vis}}$, and where have distinguished between the external and internal reflection and transmission coefficients at a given frequency, due to the asymmetry at the air-gold-quartz interfaces (see Supplementary Discussion 1, and Extended Data Figure 3). The refractive index for quartz used in this work is tabulated for reference in Extended Data Table 1.

The nonlinear polarization inside of the cavity is then computed in terms of the product of the THz and visible probing cavity fields[33,51] (i.e. using S3.1):

$$\frac{\partial}{\partial z} E_{\text{cav}}(\omega_\pm, z) \propto i \frac{\omega_\pm^2}{c_0^2 k(\omega_\pm)} \chi_{\text{eff}}^{(2)}(\Omega_{\text{THz}}) \frac{t_{\text{ext}}(\Omega_{\text{THz}}) t_{\text{ext}}(\omega_{\text{vis}}) E_{\text{inc}}^{\text{THz}}(\omega, z) E_{\text{inc}}^{\text{vis}}(\omega, z) \left( e^{i\Delta k_{\text{co}} z} + r_{\text{int}}(\Omega_{\text{THz}}) e^{-\frac{i\Omega_{\text{THz}} \tau_{\text{RT}}(\Omega_{\text{THz}})}{2}} e^{i\Delta k_{\text{xr}} z} \right)}{(1 - r_{\text{int}}(\Omega_{\text{THz}})^2 e^{-i\Omega_{\text{THz}} \tau_{\text{RT}}(\Omega_{\text{THz}})})(1 - r_{\text{int}}(\omega_{\text{vis}})^2 e^{i\omega_{\text{vis}} \tau_{\text{RT}}(\omega_{\text{vis}})})}, \tag{S3.2}$$

where $\omega_\pm = \omega_{\text{vis}} \pm \Omega_{\text{THz}}$ is the angular frequency of the emitted (SFG/DFG) field. Note that the phase in the Fabry-Pérot transfer function for the probe field is conjugated compared to the THz transfer function – a consequence of the relative time delay definition between the THz and probe pulses. We have identified the co- and counter-propagating momentum mismatch as:

$$\Delta k_{\text{co}} = \Delta k_{\text{co}}(\omega_\pm; \Omega_{\text{THz}}, \omega_{\text{vis}}) = k(\omega_\pm) - k(\omega_{\text{vis}}) \mp k(\Omega_{\text{THz}}) \approx \pm (n_{\text{THz}} - n_{\text{vis}}) \Omega_{\text{THz}},$$

$$\Delta k_{\text{xr}} = \Delta k_{\text{xr}}(\omega_\pm; \Omega_{\text{THz}}, \omega_{\text{vis}}) = k(\omega_\pm) + k(\omega_{\text{vis}}) \mp k(\Omega_{\text{THz}}) \approx \pm (n_{\text{THz}} + n_{\text{vis}}) \Omega_{\text{THz}}. \tag{S3.3}$$

We have assumed that there is negligible dispersions the refractive index in the visible frequency in order to evaluate for the final expressions on the right-hand side of Equation S3.3.

In principle, we have in total four terms to consider, given the possible combinations of THz and probing pulse propagation directions. We have assumed already in writing Equation S3.1 that the reflection of the probe inside the cavity is quite minor, eliminating the two terms arising from the backwards-travelling probing pulse. This is a relatively crude approximation, but bolstered by the fact that the probe must internally reflect twice to be finally measured in the transmission region, producing at that point a relatively very weak signal. A related approximation we make is that the SFG/DFG signal does not have the cavity reflectivity imposed upon it. Nevertheless, we observe that the principal single-cycle pulse extracted without including cross-propagation (Extended Data Fig. 2d), or including it (Supplementary Figure 5 below) suggest that this level of treatment is wholly sufficient for the moderate cavity quality factors investigated here.

The functional form of the cavity correction function displayed in the Methods is derived from a simplified form of Equation S3.1, where we do not consider the backwards-travelling THz pulse when considering the nonlinear polarization inside of the cavity. We present here a more comprehensive correction function.

$$h_{\text{fun}}(\Omega_{\text{THz}}) = \chi_{\text{eff}}^{(2)}(\Omega_{\text{THz}}) t_F(\Omega_{\text{THz}}) \int_{\omega_{\text{vis}}}^{\square} \frac{\omega_{\text{vis}}^2}{c_0^2 k(\omega_{\text{vis}})} T_{\text{pr}}(\omega_{\text{vis}}, \Omega_{\text{THz}}) E_{\text{pr}}^*(\omega_{\text{vis}}) E_{\text{pr}}(\omega_{\text{vis}} \mp \Omega_{\text{THz}}) \cdots$$

$$\left( G_{\text{co}}(\omega_{\text{vis}}, \Omega_{\text{THz}}) + r_{\text{int}}(\Omega_{\text{THz}}) e^{-\frac{i\Omega_{\text{THz}} \tau_{\text{RT}}(\Omega_{\text{THz}})}{2}} G_{\text{xr}}(\omega_{\text{vis}}, \Omega_{\text{THz}}) \right) d\omega_{\text{vis}} \tag{S3.4}$$

where the new overall phase mismatch term is distributed into the conventional one, $G_{\text{co}}(\omega_{\text{vis}}, \Omega_{\text{THz}})$ defined as in the Methods, as well as a new one corresponding to counter-propagating THz and probing pulses:

$$G_{\text{xr}}(\omega_{\text{vis}}, \Omega_{\text{THz}}; L_{\text{Qtz}}) = \left( \frac{e^{i\Delta k_{xr}(\Omega_{\text{THz}}, \omega_{\text{vis}}) L_{\text{Qtz}}} - 1}{i\Delta k_{co}(\Omega_{\text{THz}}, \omega_{\text{vis}})} \right) \tag{S3.5}$$

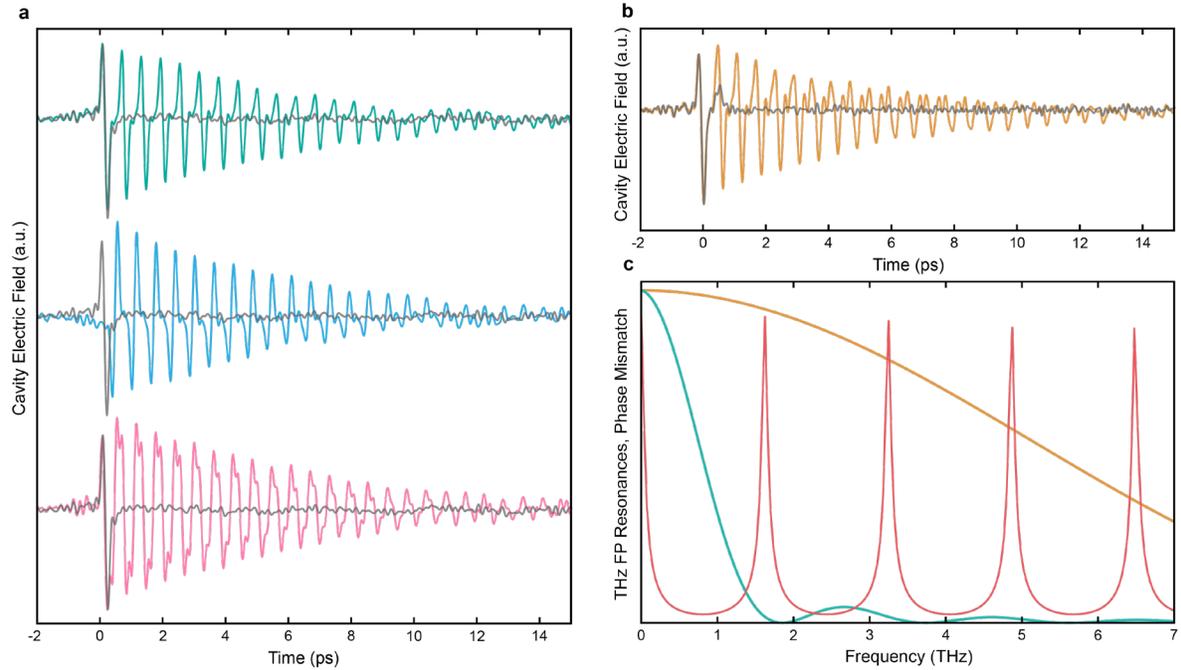

**Supplementary Fig. 5| Full (Cross-Propagating) Cavity Correction Function. a**, The forward-propagating field (green, top), backwards-propagating field (blue, middle), and the sum (pink, bottom), evaluated at the cavity center (z=0), using the principal pulse (gray) obtained from the full cavity correction function. **b**, The principal pulse (gray) and cavity field (orange) obtained using the approximate form of the cavity correction function, where weak detection of the counter-propagating THz pulse is still present. **c**, The Fabry-Pérot resonances (red; $L_{QTz}$ = 44 µm) are generally near the zeros of the counter-propagating phase mismatch function (green), plotted here in comparison with the co-propagating phase mismatch function (orange).

We show the convolution of this field with the backwards-propagating cavity field transfer function (evaluated at $z$ = 0). We note that this is an expected field, rather than a direct measurement, as we only very weakly measure the backwards-propagating field, as expected from the severe phase mismatch (Supp. Fig. 5c), and evidenced by the relatively minor difference between the extracted principal pulses with and without counter-propagating fields (Supp. Figs. 5a,b, respectively). That is, we display the total field for purposes of demonstrating the expected cavity fields at the center of the cavity, despite electro-optic sampling primarily measuring co-propagating pulses.

## Supplementary Discussion 4
Cavity Electro-Optic Sampling – Role of Phase Mismatch in Experimental EOC Spectra

We note that the effects of phase mismatch as it is defined in Methods equation 3 are responsible for noise injected (i.e. division by zero) into the spectrum from the cavity correction function. We depict below the normalized phase mismatch function as a function of quartz crystal thickness, where we identify the first zero-crossing, and reproduce this on the experimental cavity EO spectra shown in **Fig. 2b**. The normalized phase mismatch function is defined as:

$$G(\omega_{\text{vis}}, \Omega_{\text{THz}}; L_{\text{Qtz}}) = \frac{e^{i\Delta k_{\text{co}}(\Omega_{\text{THz}},\omega_{\text{vis}})L_{\text{Qtz}}} - 1}{i\Delta k_{\text{co}}(\Omega_{\text{THz}},\omega_{\text{vis}})L_{\text{Qtz}}}, \tag{S4.1}$$

where we have used the definition of the co-propagating momentum mismatch as written in Equation S3.3.

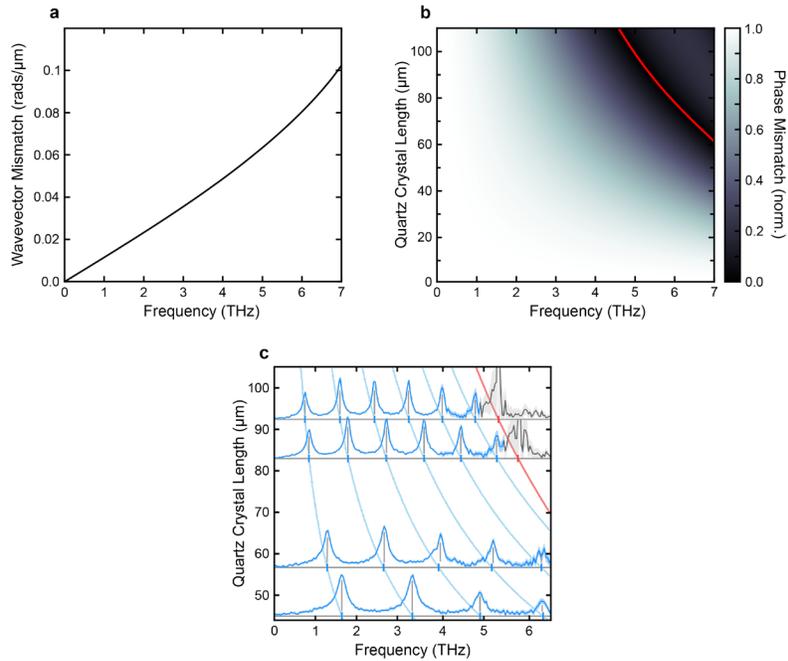

**Supplementary Fig. 6| Effect of Phase Mismatch Function. a,** The momentum mismatch is displayed as a function of THz refractive index. **b,** The (normalized) momentum mismatch function is displayed as a function of both THz frequency and quartz crystal length $L_{Qtz}$. The first zero of this phase mismatch function is plotted here as a red line. **c,** The EOC spectrum as a function of quartz length is repeated here from **Fig. 2b**, where we additionally overlay the first zero of the phase mismatch function, to highlight the origin of the injected noise which we exempt from our time-domain fields by application of a masking function.

## Supplementary Discussion 5
Probe Scattering – Field Calibration

For the purposes of measuring the electric field strength quantitatively, we need to know the field strengths of the probe laser inside of the electro-optic medium. This is typically achieved by measuring the intensity of the probe laser after the sample, and then using the refractive index of the electro-optic crystal to infer the field strength inside the crystal. In our case, however, there is scattering and near-field optical effects with the probe laser and the gold islands that will have to be considered in order to properly calibrate the probe fields which give rise to the cavity electro-optic signal.

We assume that there are losses to the probe beam intensity, whether due to scattering or reflection, which are proportional to the incident intensity, leading to the following equations for the intensity inside the EOC and in the transmission region, for a given nominal gold film thickness $d_{Au}$:

$$\mathrm{I}_{EOC}(d_{Au}) = t_{Au}(d_{Au}) t_{ext}(d_{Au}=0) I_o, \qquad \mathrm{I}_{trans.} = t_{Au}^2(d_{Au})\, t_{int}(d_{Au}=0)\, t_{ext}(d_{Au}=0)\, \mathrm{I}_o. \tag{S5.1}$$

Here we have called the transmission through the gold islands as $t_{Au}$. We assume that the entire optical response, prior to the establishment of gold films is attributed to scattering, or generally near-field, or nano-optical effects, of the probe with gold islands, such that by comparing the measured probe intensities as a function of nominal gold film thickness we can infer the transmission through a single gold film:

$$t_{Au}(d_{Au}) = \sqrt{\frac{I_{trans}(d_{Au})}{\mathrm{I}_{EOC}(0)}}. \tag{S5.2}$$

We compare our experimentally-inferred gold film transmittance values using the above expression, with numerical simulations[6], where we observe very good agreement. The general trend is understood as a combination of three effects: the evolution of the scattering cross-section as the gold islands morphologies develops with nominal film thickness, the reflectivity from the partially-established gold films, and finally the evolution of the gold plasmon frequency dramatically red-shifting across the spectrum of the probe polarization. The competition for these effects leads to the highly non-monotonic curve displayed in both our experimental data and the theoretical data in Supplementary Figure 7. We use these experimental transmission values to correct the field strengths that we plot in **Fig. 2e**

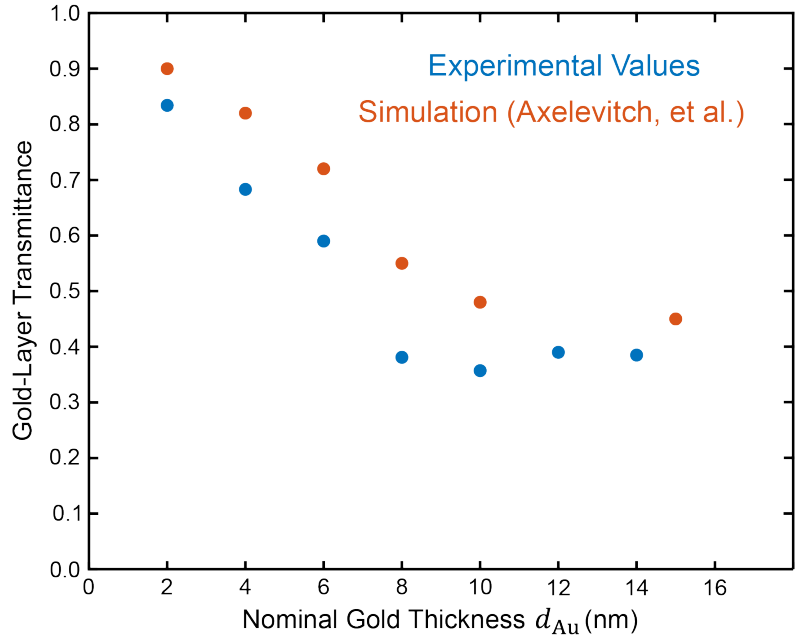

**Supplementary Fig. 7| Gold Island Transmission.** We compare our experimentally-inferred gold film transmittance with simulation provided in Alexevitch et al.[6], where we have estimated the values at approximately 800 nm from their results.

## Supplementary Discussion 6
Cavity Field Model

We identify the spatial functional form of the cavity electric fields by solving for the cavity eigenmodes. We consider a plane wave in each region (see Extended Data Figure 4), and impose the following boundary conditions: (1) we search for cavity modes, i.e. modes whose amplitudes go to zero at the gold interface, (2) fields which are continuous across the interfaces, and (3) fields which have a continuous derivative.

By forcing the field to go to zero at the gold boundaries (i.e. $z = \pm(L_o + L_{Qtz})$) we then write the fields in the left and right quartz as:

$$E_{Qtz}^{(L)}(z) = A\left(e^{ik_{Qtz}z} - e^{-2ik_{Qtz}(L_o+L_{Qtz})}e^{-ik_{Qtz}z}\right) \tag{S6.1}$$

$$E_{Qtz}^{(R)}(z) = F\left(e^{ik_{Qtz}z} - e^{2ik_{Qtz}(L_o+L_{Qtz})}e^{-ik_{Qtz}z}\right) \tag{S6.2}$$

We next enforce continuity across the air-quartz interfaces:

$$E_{Qtz}^{(L)}(z)\Big|_{z=-L_o} = E_{Air}(z)|_{z=-L_o} \rightarrow 2iAe^{-ik_{Qtz}(L_o+L_{Qtz})}\sin(k_{Qtz}L_{Qtz}) = Ce^{-ik_oL_o} + De^{ik_oL_o} \tag{S6.3}$$

$$E_{Qtz}^{(R)}(z)\Big|_{z=L_o} = E_{Air}(z)|_{z=L_o} \rightarrow -2iFe^{ik_{Qtz}(L_o+L_{Qtz})}\sin(k_{Qtz}L_{Qtz}) = Ce^{ik_oL_o} + De^{-ik_oL_o} \tag{S6.4}$$

Finally, we can enforce continuity of the first derivative:

$$\frac{\partial E_{Qtz}^{(L)}(z)}{\partial z}\bigg|_{z=-L_o} = \frac{\partial E_{Air}(z)}{\partial z}\bigg|_{z=-L_o} \rightarrow \frac{2k_{Qtz}}{k_o}Ae^{-ik_{Qtz}(L_o+L_{Qtz})}\cos(k_{Qtz}L_{Qtz}) = Ce^{-ik_oL_o} - De^{ik_oL_o} \tag{S6.5}$$

$$\frac{\partial E_{Qtz}^{(R)}(z)}{\partial z}\bigg|_{z=L_o} = \frac{\partial E_{Air}(z)}{\partial z}\bigg|_{z=L_o} \rightarrow \frac{2k_{Qtz}}{k_o}Fe^{ik_{Qtz}(L_o+L_{Qtz})}\cos(k_{Qtz}L_{Qtz}) = Ce^{ik_oL_o} + De^{-ik_oL_o} \tag{S6.6}$$

With these last four equations, we want to identify solutions which satisfy all the equations. By repeated substitution, we identify the following two expressions, which are the transcendental equations that can produce the possible eigenvalues.

$$\frac{k_{Qtz}}{k_o}\tan(k_o L_o) = -\tan(k_{Qtz}L_{Qtz}); \quad \frac{k_{Qtz}}{k_o}\cot(k_o L_o) = \tan(k_{Qtz}L_{Qtz}) \tag{S6.7}$$

Here the first equation holds only when $C - D \neq 0$, and similarly the second only when $C + D \neq 0$. We identify these equations then as physically referring to odd ($C = -D$) and even ($C = D$) modes, respectively. By using either of these restrictions on the relation between forward- and backwards-propagating plane waves in air, it is straightforward to obtain the electric field defined in Methods from equations S4.3, and S4.4.

# References

1. Ismail, N., Kores, C. C., Geskus, D. & Pollnau, M. Fabry-Pérot resonator: spectral line shapes, generic and related Airy distributions, linewidths, finesses, and performance at low or frequency-dependent reflectivity. *Opt. Express* 24, 16366 (2016).
2. Frenzel, M. *et al.* Quartz as an Accurate High-Field Low-Cost THz Helicity Detector. (2023).
3. Kampfrath, T., Nötzold, J. & Wolf, M. Sampling of broadband terahertz pulses with thick electro-optic crystals. *Appl. Phys. Lett.* 90, 231113 (2007).
4. Yuen-Ron Shen. *The Principles of Nonlinear Optics*. (Wiley, 1984).
5. Gallot, G. & Grischkowsky, D. Electro-optic detection of terahertz radiation. *Journal of the Optical Society of America B* 16, 1204 (1999).
6. Axelevitch, A., Apter, B. & Golan, G. Simulation and experimental investigation of optical transparency in gold island films. *Opt. Express* **21**, 4126 (2013).